\documentclass[reprint,superscriptaddress]{revtex4-1}
\usepackage[utf8]{inputenc}
\usepackage{amsfonts}
\usepackage{amssymb}
\usepackage{amsmath}
\usepackage{bbm}
\usepackage{verbatim}
\usepackage{graphicx}
\usepackage{xcolor}
\usepackage{hyperref}
\usepackage{listings}
\hypersetup{
    colorlinks = true,
    citecolor = {blue},
}

\usepackage{setspace}

\newcommand{\imag}{\mathrm{Im}\,}
\newcommand{\real}{\mathrm{Re}\,}

\newcommand{\R}{\mathbb{R}}
\newcommand{\C}{\mathbb{C}}

\newcommand{\sign}{\mathrm{sign}}

\newcommand{\dP}{\mathrm{d}P}
\newcommand{\dU}{\mathrm{d}U}

\newcommand{\trace}{\mathrm{Tr}}

\newcommand{\SUThree}{\mathrm{SU}(3)}
\newcommand{\SUN}{\mathrm{SU}(N)}

\newcommand{\UN}{\mathrm{U}(N)}

\newcommand{\UOne}{\mathrm{U}(1)}

\begin{document}
\title{Simulating Yang-Mills theories with a complex coupling}

\author{Jan M. Pawlowski}
\affiliation{Institut f\"ur Theoretische Physik, Universit\"at Heidelberg, Philosophenweg 16, D-69120 Heidelberg, Germany}
\affiliation{ExtreMe Matter Institute EMMI, GSI, Planckstra\ss e 1, D-64291 Darmstadt, Germany}

\author{Manuel Scherzer}
\affiliation{Institut f\"ur Theoretische Physik, Universit\"at Heidelberg, Philosophenweg 16, D-69120 Heidelberg, Germany}

\author{Christian Schmidt} 
\affiliation{Fakult\"at f\"ur Physik, Universit\"at Bielefeld, Postfach 100131, D-33501 Bielefeld, Germany.}

\author{Felix P.G. Ziegler}
\affiliation{CP3-Origins and Danish IAS,
Department of Mathematics and Computer Science, University of Southern Denmark, Campusvej 55, 5230 Odense M, Denmark}
\author{Felix Ziesch\'e}
\affiliation{Fakult\"at f\"ur Physik, Universit\"at Bielefeld, Postfach 100131, D-33501 Bielefeld, Germany.}
\date{\today}
\begin{abstract}
We propose a novel simulation strategy for Yang-Mills theories with a complex  coupling, based on the Lefschetz thimble decomposition. We envisage, that the approach developed in the present work, can also be adapted to QCD at finite density, and real time simulations. 

Simulations with Lefschetz thimbles offer a potential solution to sign problems in Monte Carlo calculations within many different models with complex actions. We discuss the structure of Generalized Lefschetz thimbles for pure Yang-Mills theories with a complex gauge coupling $\beta$ and show how to incorporate the gauge orbits. We propose to simulate such theories on the union of the tangential manifolds to the relevant Lefschetz thimbles attached to the critical manifolds of the Yang-Mills action. We demonstrate our algorithm on a
(1+1)-dimensional U(1) model and discuss how, starting from the main thimble result, successive subleading thimbles can be taken into account via a reweighting approach. While we face a residual sign problem, our novel approach performs exponentially better than the standard reweighting approach. 
\end{abstract}

\maketitle

\section{Introduction}

The notorious sign problem hampers numerical simulations of many interesting 
physical systems, ranging from high energy to condensed matter systems. A sign 
problem is faced in numerical calculations of statistical models whenever the 
action becomes genuinely complex. Hence, standard Monte Carlo methods and in 
particular importance sampling drastically lose their efficiency with increasing lattice volume.

Examples of theories with a sign problem include real time calculations in lattice-regularized quantum field theories, i.e.~lattice QCD in Minkowski space-time, lattice QCD with a nonzero vacuum angle $\theta$, and lattice QCD with a nonzero baryon chemical 
potential $\mu_B$. For the latter two cases, many methods have been developed that
potentially circumvent or solve this problem in the continuum limit. These methods include reweighting \cite{Barbour:1997ej, Fodor:2001au}, 
Taylor expansions \cite{Allton:2002zi, Allton:2003vx, Gavai:2003mf}, analytic 
continuation from 
purely imaginary chemical potentials \cite{deForcrand:2002hgr, DElia:2002tig}, 
canonical partition functions \cite{Kratochvila:2003rp, Alexandru:2005ix}, 
strong coupling/dual methods \cite{Karsch:1988zx, deForcrand:2014tha, Gattringer:2016kco, Gagliardi:2019cpa},
the density of states method \cite{Ambjorn:2002pz, Fodor:2007vv, Langfeld:2015fua}, and complex Langevin dynamics \cite{Karsch:1985cb, Aarts:2009uq, Aarts:2012ft, Sexty:2013ica}, see \cite{Attanasio:2020spv} for a review of recent
developments. 
However, all these methods so far face severe limitations that restrict their applicability in the continuum limit. 

In the past decade deformations of the original integration manifold into the complex domain have been introduced, based on complex saddle points of the action, the Lefschetz thimbles \cite{Pham, Witten:2010zr}. If these  deformations are chosen well, all physical expectation values obtained from an oscillatory integral remain unchanged but the sign problem is drastically alleviated. Thus, a numerical evaluation of the theory is accessible. By definition, the imaginary part of the action (phase of the probability density) is stationary on the thimble. A first Lefschetz thimble algorithm in the context of the QCD finite density sign problem was introduced in \cite{Cristoforetti:2012su}, for a recent review see \cite{Alexandru:2020wrj}. Despite its great potential for beating the sign problem, simulations with Lefschetz thimbles have to overcome the following intricacies: 
\begin{itemize}
\item[($i$)] 
A parametrization of the thimble is a priori unknown and has to
be obtained as the numerical solution of a flow-equation. The parametrization and the necessary Jacobian of the variable transformation are numerically demanding.
\item[($ii$)] The 
curvature of the thimble manifold introduces a so-called residual sign problem through the Jacobian. 
\item[($iii$)] In most cases, one has to include relevant contributions from a large number of thimbles. Their relative weight may give rise to a further, residual, sign problem. In any case, the probability density becomes multi-modal.
\end{itemize}
These shortcomings have been addressed by formulating particularly efficient methods for the
calculation of the Jacobian \cite{Cristoforetti:2014gsa, Alexandru:2017lqr}, by optimizing
the deformation of the integration domain either based on a model ansatz \cite{Bursa:2018ykf,
Mori:2017pne} or by means of machine learning \cite{Mori:2017nwj, Alexandru:2017czx,
Wynen:2020uzx} and finally by using tempering \cite{Alexandru:2017oyw, Fukuma:2019wbv} and
other advanced strategies to foster transitions between thimbles and take into account
contributions from multiple thimbles \cite{DiRenzo:2017igr, Bluecher:2018sgj,
DiRenzo:2020cgp}.

In the present work we analyze the structure of the Lefschetz thimble decomposition of pure gauge theories
with gauge groups $\text{U}(N)$ and $\text{SU}(N)$ and complex coupling $\beta$. We propose
to sample on the tangential manifold attached to the thimble of the main critical point,
\textit{i.e.} the critical point with the smallest action value. As expected, we find a
full hierarchy of critical points (saddle points) that have to be considered depending on
the coupling parameter $\beta$. We include successive subleading saddle point contributions
via reweighting. The reweighting procedure is set up by using linear mappings from the main
tangential manifold to the tangential manifold attached to the thimble of the target
critical point. 

\begin{figure}[t]
\includegraphics[height=6cm]{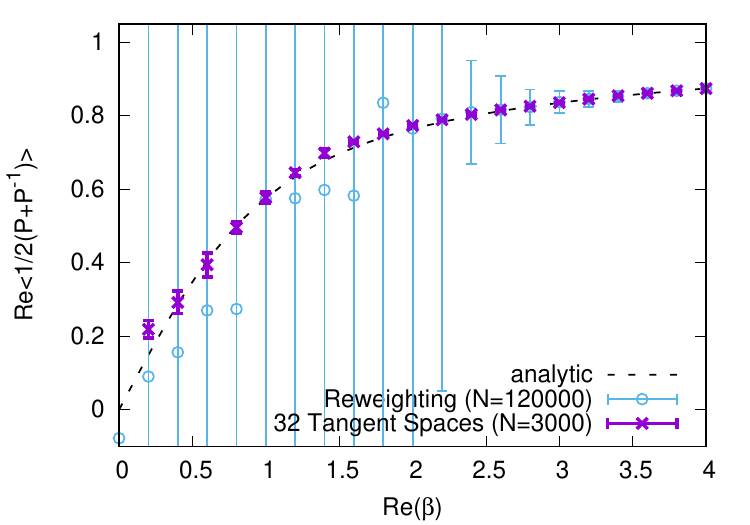}
\caption{Comparison of the novel Takagi simulation method with standard reweighting on
an $8 \times 8$ lattice with gauge group $\UOne$, where we measured the real part of the
average plaquette. We vary the real part of $\beta$, while keeping
$\imag (\beta) = 1$ constant.
Both reweighting and our approach took about the same amount of computing time.}
\label{fig:U1result}
\end{figure}

Our approach has the following advantages over the flow-based generalized
Lefschetz thimble approach \cite{Alexandru:2015xva}: 
\begin{itemize}
    \item[($A1$)] 
    During the sampling procedure,
we neither have to flow our configuration nor have to calculate a Jacobian since it is
constant on the tangential manifold.
\item[($A2$)] As contributions from subleading thimbles are
taken into account by reweighting \cite{Bluecher:2018sgj}, there is no ergodicity problem
due to potential barriers between thimbles. The reweighting procedure does also not
introduce any overlap problem since critical points are mapped onto critical points.
\end{itemize}   
The disadvantages are: 
\begin{itemize}
    \item[($D1$)] Sampling on the tangential manifold rather than on the thimble
itself does not completely resolve the sign problem, however, there is no residual sign
problem from the Jacobian \cite{Cristoforetti:2014gsa}. 
\item[($D2$)] 
The critical points need to
be known.  
\end{itemize}
The advantages are clearly demonstrated for the benchmark case of an $\UOne$ gauge theory, as summarized in Fig.~\ref{fig:U1result}. There we compare our results for the real part of the plaquette with that produced by the standard reweighting procedure. For the same numerical costs the error bars of our results are reduced by orders of magnitude. 

This will be discussed in more detail in this work, which is organized as follows: In Sec.~\ref{sec:formulation} we derive necessary
formulas for our setup. Emphasis is devoted to discussing 
the critical manifolds of the Yang-Mills action as well as the Lefschetz thimbles. Moreover we explain our update and
reweighting procedures. In Sec.~\ref{sec:application} we apply our algorithm to the case
of a 2-dimensional $\text{U}(1)$ gauge theory. In Sec.~\ref{sec:discussion} we discuss prospect of our approach. 
Finally we conclude in Sec.~\ref{sec:conclusion}. 

\section{Formulation \label{sec:formulation}}
\subsection{Overview}
We consider the standard discretization of the Yang-Mills action 
\cite{Wilson:1974sk}
\begin{equation}
  S = \beta\sum_{x}\sum_{\mu<\nu}\left\{1-\frac{1}{2N}
  \left(\text{Tr}P_{\mu,\nu}(x)+\text{Tr}P^{-1}_{\mu,\nu}(x)\right)\right\}\,,
\label{wilson_action}
\end{equation}
where
$P_{\mu,\nu}(x)=U_\mu(x)U_\nu(x+\hat\mu)U^{-1}_\mu(x+\hat\nu)U^{-1}_\nu(x)$
denotes the elementary plaquette in the ($\mu,\nu$)-plane at lattice site
$x \in \Lambda$ where $\Lambda \subset \mathbb{Z}^d$. The summation is done such that each plaquette is counted with
only one orientation. The link variables $U_\mu(x)$ are elements of
the gauge group, which we consider to be a Lie group and in particular 
$\text{U}(N)$ or $\text{SU}(N)$.

A sign problem is introduced by choosing general complex couplings
$\beta$.

We aim to update a given
gauge field configuration such that the imaginary part of the action
$\text{Im}(S)$ varies slowly and hence
the remaining sign problem is mild. 

In order to
achieve this goal the link variables are 
complexified, {\it i.e.}~they are allowed to take values 
within the larger group
$\text{GL}(N,\mathbb{C})$ or $\text{SL}(N,\mathbb{C})$,
respectively. 
By generalized Picard-Lefschetz theory, there is a smooth
middle dimensional manifold connected to each complex 
critical manifold (simply connected union of points, where $\nabla S = 0$)  of
the action on which $\text{Im}(S)$ stays constant. These manifolds are
called Lefschetz thimbles. As already stated, updates that stay on
these thimbles are computational very demanding and are invariably global updates 
\cite{Cristoforetti:2012su, Cristoforetti:2013wha,
  Fujii:2013sra, Alexandru:2015xva}. 
To reduce the computational
demand, it has been argued that 
one does not have to stay exactly on the thimble 
in order to reduce the sign problem 
significantly \cite{Alexandru:2015sua, Mori:2017pne}. 
Any deformation of the original integration domain
which has the correct asymptotic behavior will be sufficient. 
Here we construct 
an integral deformation, which is the union of all
tangential manifolds to the critical points. 
As for the pure gauge
theory all critical manifolds are known, 
this manifold is straightforward
to parametrize. 
After discussing the critical manifolds in Sec.~\ref{sec:ym-crit-1} and \ref{sec:ym-crit-2}
we discuss properties of Lefschetz thimbles and tangential manifolds
in Sec.~\ref{sec:lefschetz-and-tang} and \ref{sec:hierarchy}. Our algorithmic 
approach is presented in Sec.~\ref{sec:GenTakagiUpdates}, 
\ref{sec:AltUpdates} and \ref{sec:algo-subl}.

\subsection{The critical points of the Yang-Mills action}
\label{sec:ym-crit-1}
A Lefschetz thimble is generally defined 
to be the union of flow lines
generated by the steepest descent equation
\begin{equation}
\frac{\mathrm{d}U}{\mathrm{d}t} 
= -\left(\frac{\delta S}{\delta U}\right)^*
\,,
\label{SteepestDescent}
\end{equation}
which end in a non-degenerate critical point 
of the action. The degeneracy of critical points due to gauge symmetry necessitates the application of Witten's concept of generalized Lefschetz thimbles \cite[3.3]{Witten:2010cx}.

The critical manifolds can be described 
in terms of plaquette variables.
This is seen by examining the gradient of the action
\begin{align}
&\,\partial_{x, \kappa, a} S = 
-\frac{i\beta}{2N}\times \label{CritEq} \\[2ex]
&\,\trace\left[\left(\sum_{\kappa < \nu}
P_{\kappa,\nu}(x)-P^{-1}_{\kappa,\nu}(x)+P_{\kappa,-\nu}(x)-P^{-1}_{\kappa,-\nu}(x)
\right.\right. \nonumber \\[2ex]
&\,- \left.\left.\sum_{\mu < \kappa}
P_{\mu,\kappa}(x)-P^{-1}_{\mu,\kappa}(x)+P_{-\mu,\kappa}(x)-P^{-1}_{-\mu,\kappa}(x)
\right)t_a\right]\,.\nonumber
\end{align}
Here $t_a$ are the generator matrices of
the related Lie-Algebras $\mathfrak{su}(N),\mathfrak{u}(N)$.
In our notation those are Hermitian, and 
for $N>1$ they satisfy the 
normalization condition $\trace[t_a t_b]=\frac{1}{2}\delta_{ab}$.
The derivative with respect to the gauge link $U$ in the 
direction of $t_a$ is 
defined as $\partial_{a} f(U) := \frac{\partial}{\partial \omega}
f(e^{i \omega t_a}U)|_{\omega = 0}$.
Negative signs in the subscript of the plaquette variables refer to reversed 
directions in their orientation, e.g.
\begin{equation*}
P_{\kappa,-\nu}(x) = U_{\kappa}(x)U_{\nu}^{-1}(x + \hat\kappa - \hat\nu)U_{\kappa}^{-1}(x-\hat\nu)U_{\nu}(x-\hat\nu)\,.
\label{eq:ex-rev-plaq}
\end{equation*}
A necessary condition for a critical 
configuration is a vanishing  gradient of the action.

In the following we derive relations that constrain 
possible plaquette values from a critical
configuration. 
Eq.~(\ref{CritEq}) vanishes $\forall \ a$, if the matrix in round brackets is
proportional to $\mathbbm{1}$
for plaquette values in $\text{SL}(N,\C)$.
For plaquette values in $\text{GL}(N,\C)$, the matrix has to be zero.
For a proof, see the App.~\ref{midentity}.
This criticality condition yields
relations for adjacent plaquettes sharing one link. 
Note that in $d$ dimension one 
link is shared by $2(d-1)$ plaquettes. 

We exemplify the $d=2$ case: For plaquette values
in $\text{GL}(N,\C)$, we can directly read off the relations

\begin{equation}
P_{1,2}(x) = P^{-1}_{1,-2}(x)\:\: \text{or}\:\: P_{1,2}(x) = -P_{1,-2}(x)\, ,\nonumber
\end{equation}
and
\begin{equation}
P_{1,2}(x) = P^{-1}_{-1,2}(x)\:\: \text{or}\:\: P_{1,2}(x) = -P_{-1,2}(x)\;.
\label{CritUN}
\end{equation}
If we assume that our critical configuration consists of commuting (Abelian) link variables, \textit{i.e.} if link variables are diagonal, the above relation simplifies to
\begin{equation}
P_{1,2}(x) = P_{1,2}(x-\hat\nu) \;\;\text{or}\;\; P_{1,2}(x) = -P^{-1}_{1,2}(x-\hat\nu)\,,
\end{equation}
with $\nu\in\{1,2\}$.
It follows that each critical configuration 
exhibits at most two distinct plaquettes values.
If plaquettes take values in $\text{SL}(N,\C)$ we get, \textit{e.g.}, for a link in $\hat{1}$-direction
\begin{equation}
(P_{1,2}(x)-P^{-1}_{1,2}(x)-P^{-1}_{1,-2}(x)+P_{1,-2}(x)) = \alpha\mathbbm{1}\,,
\label{CritSUN}
\end{equation}
for an arbitrary $\alpha\in\C$.
For the Abelian case, this equation reformulates again to a relation between adjacent plaquettes. Restricting this further to the original group $\SUN$, \textit{i.e.} the maximal torus of $\SUN$, we find
\begin{multline}
(\imag P_{1,2}(x) - \imag P_{1,2}(x-\hat\nu))_{ii} = \\
(\imag P_{1,2}(x) - \imag P_{1,2}(x-\hat\nu))_{jj}
\;\forall i,j,\nu\,.
\label{eq:crit-sun-imag}
\end{multline}
For $d >2$, we obtain the same constraint on the imaginary parts of diagonal entries of adjacent plaquettes, but naturally there are more adjacent plaquettes.

From the periodic boundary condition we can derive further constraints. Under the assumption, that the relevant critical configurations are Abelian, the product of
all plaquettes in an arbitrary 2-dimensional
hyperplane $\Lambda_{\mu\nu}$ must be one, \textit{i.e.}
\begin{equation}
\prod_{x\in\Lambda_{\mu\nu}} P_{\mu\nu}(x) = 1\,.
\label{PerCon}
\end{equation}

\subsection{Locality and critical manifolds}
\label{sec:ym-crit-2}

We are left with a fairly large number of critical configurations,
which we will boil down to a set of basis configurations, getting
the others by transpositions and symmetry relations.
The ultimate aim is to obtain a homotopic covering 
of the original integration
domain $[\UN]^{dV}$ or $[\SUN]^{dV}$\,.\\

As a guiding inspiration, we start with a discussion of the one-plaquette model, 
\textit{i.e.} we just have only one plaquette degree of freedom, as Eq.~(\ref{wilson_action}) is a sum of local plaquette terms.  The action is defined 
as
\begin{equation}
S = -\frac{\beta}{2N}\trace\left[P+P^{-1}\right]\,,
\label{OnePlaqAct}
\end{equation}
where an irrelevant constant has been
omitted.
The Lie derivatives with respect to $P$ are given by 
\begin{equation}
\partial_a S = -\frac{i\beta}{2N}\trace
\left[(P-P^{-1})t_a\right]\,.
\label{OnePlaqActDerv}
\end{equation}
Observations are:
\begin{itemize}
    \item[(i)] Eq. (\ref{OnePlaqActDerv}) 
    vanishes for self-inverse plaquettes $P$ 
    in $\UN$. 
    Therefore, the critical points
    consists only of matrices 
    whose eigenvalues are $+1$ and $-1$.
    \item[(ii)] For $P\in\SUN$ Eq.~(\ref{eq:crit-sun-imag}) implies
    that the imaginary part of all eigenvalues must be identical. For vanishing imaginary part we obtain 
    the self-inverse elements of $\SUN$. The constraint $\det[P]=1$ implies  
    an even number of ($-1$)-eigenvalues, denoted by $N^{(-)}$. For non-vanishing imaginary part,
    the center elements of $\SUN$ are solutions.
    For $N\geq6$, we
    have roots of unity apart from $\pm 1$ with the same imaginary parts. So a mixing of
    these yielding a unit determinant is a valid solution.
\end{itemize}
Different critical points indicate different action values. Their importance with 
respect to the weight factor $w\equiv e^{-\real(S)}$ may be exponentially suppressed. 
For the one-plaquette model we find the following 
hierarchy of critical points: For $P\in\UN$, we have
\begin{equation}
S = -\frac{\beta}{N}\left(N - 
N^{(-)}\right)\,.
\end{equation}
The importance of a critical point thus shrinks with the number of
($-1$)-eigenvalues. 
For $\SUThree$, we find 6
critical points with three different action values
\begin{align}
    S = -\beta\,, &\quad P = \mathbbm{1}
    \nonumber \\[1ex]
    S = -\frac{\beta}{2}\,, &\quad  P \in\left\{ e^{i\frac{2\pi}{3}k}\mathbbm{1}\,,\; k=1,2 \right\}
    \nonumber \\[1ex]
    S = -\frac{\beta}{3}, &\quad  P\in\left\{
    \text{diag}(1,-1,-1),\; \text{diag}(-1,1,-1),\right.
    \nonumber \\[1ex]
    & \left.\qquad \quad \text{diag}(-1,-1,1)
    \right\}\,.
\end{align}

Critical points from the one-plaquette model can 
be used to construct certain critical 
configurations for the full lattice theory. 
We pick critical plaquette values from the
one-plaquette model and distribute 
them in accordance with Eq.~(\ref{CritUN}) and
(\ref{CritSUN}) over the lattice. 
Each critical configuration obtained in this way 
is one representative of a critical manifold, 
which consists of its gauge orbit and additional 
zero modes of the action. For this simple procedure 
we obtain an additional constraint from
Eq.~(\ref{PerCon}): The product of eigenvalues at 
every position over every two-dimensional
hyperplane must be one. We are therefore 
limited to configurations, where the number 
of ($-1$)-eigenvalues at a given diagonal entry is even (the $\UN$ case)
in every hyperplane or in principal their
overall product is one including additional roots of unity.

Note however, that not all
critical configurations can be found by the above prescription.
Recall that Eq.~(\ref{CritUN}) restricts the plaquette values in a hyperplane at any given
position to only two possible values. We can construct a further critical configuration
by setting one (or more) plaquette values in that hyperplane to $e^{i\epsilon}$, while
choosing $e^{i(\pi - \epsilon)}$ for all remaining plaquettes in the hyperplane.
Possible $\epsilon$ values are constrained by Eq.~(\ref{PerCon}), and hence 
\begin{equation}
    \label{eq:eps-top-eq}
    k\pi + (V-2k)\epsilon = 2\pi l\,,
\end{equation}
where $k$ is the number of plaquettes with the specific diagonal entry
$e^{i(\pi - \epsilon)}$. The $2\pi l$ factor stems from the $2\pi$ periodicity.
For $d=2$, $N=1$, $l$ is the actual topological charge. It is constant on
the thimble, since the anti-holomorphic gradient flow can be seen as an
analytic continuation of the classical gradient flow, which leaves the 
topological charge invariant \cite{Luscher:2010iy}.
Note especially for $k \neq V/2$, that $\epsilon$ has to
be a real number, so the critical configurations are all in the
original group space. Picard-Lefschetz theory tells us now, that if the
tangent space of the thimble is not normal at this point to the original
group manifold, then the intersection number is non-zero (in our case one).
We will see, that for $\imag\beta \neq 0$, this is the case and they all
contribute. For $\SUN$, we have to add the constrain that the determinant equals one, 
effectively reducing the number of critical configurations.

\subsection{The Takagi decomposition and generalized Lefschetz thimbles}
\label{sec:lefschetz-and-tang}
Next, we construct the tangent spaces 
at each critical manifold described Sec.~\ref{sec:ym-crit-2}. To that end we solve
the Takagi equation
\begin{equation}
\label{takagi}
H^*\xi^* = \lambda\xi\:\:\text{with}\:\:\lambda\in\R\,.
\end{equation}
Here $H$ denotes the Hessian of the action
evaluated on the critical manifold.
Modes $\xi$ associated with positive $\lambda$, 
point in the direction of the thimble. 
In the following we refer to such a mode 
as \emph{Takagi vector}.
Correspondingly, a mode $\xi$ associated with a 
negative $\lambda$ points in direction of the anti-thimble.
It is called \emph{anti-Takagi vector} (see e.g. \cite{Alexandru:2015xva}) The
$\lambda = 0$ vectors do not 
change the action and are therefore referred
to as zero-modes. They result 
for instance from the gauge degrees
of freedom.
The Hessian of the action can be written as
\begin{eqnarray}
 \label{eq:Hessian}
 \partial_{x,\kappa,a}\partial_{y,\eta,b} S & = &
    \frac{\beta}{2N_c}\trace\left[\left(
    \sum_{U_{\kappa}(x),U_{\eta}(y)\in P}\hspace*{-20pt}(P+P^{-1})\right.\right.\nonumber \\[2ex]
    \; & + &
    \sum_{U^{-1}_{\kappa}(x),U^{-1}_{\eta}(y)\in P}\hspace*{-20pt}(P+P^{-1})\nonumber \\[2ex]
    \; & - &
    \sum_{U_{\kappa}(x), U^{-1}_{\eta}(y)\in P}\hspace*{-20pt}(P+P^{-1}) \nonumber \\[2ex]
    \; & - & \left.\left.
    \sum_{U^{-1}_{\kappa}(x), U_{\eta}(y)\in P}
    \hspace*{-20pt}(P+P^{-1})\right)
    t_b t_a\right]\,,
\end{eqnarray}
where the plaquettes appear in different orientations depending on
the position of the referred link. Since by construction we have considered  only those representatives of our critical manifolds which have diagonal links. They are Abelian and we can permute our variables to bring all generators to the right. 

For critical configurations, where the plaquettes are elements of the
original group, the Hessian splits into a real matrix with a complex prefactor $H=\beta M$, 
whose eigenvectors $v$ and eigenvalues $\alpha$ can be computed. For $\alpha\neq 0$, the solutions to Eq.~(\ref{takagi}) take the form
\begin{equation}
\label{Takagi}
\xi^{(\pm)} = \sqrt{\frac{\pm\sign(\alpha)\beta^*}{|\beta |}}\;v\,,
\end{equation}
where the $\xi^{(+)}$ ($\xi^{(-)}$) indicate the thimble (anti-thimble) 
directions. The $\alpha = 0$ eigenvectors correspond to the $\lambda = 0$ 
solutions of Eq.~(\ref{takagi}) and can have an arbitrary complex
prefactor. 

We observe that for most of our critical configurations. 
We will come to the exceptions
in Sec.~\ref{sec:U1-crit-hierarchy}.
The Hessian does not change under field
transformations in the direction of these zero modes. Consequently the
critical manifold
$\{U^{\mathrm{crit},0}_\mu(x)\}$ is independent of
those.
Therefore, we can deduce that the projection of the subspace
spanned by its zero modes in the Lie algebra is
the critical manifold itself. 
\begin{multline}
\left\{
U^{\mathrm{crit}}_\mu(x) = U^{\mathrm{crit},0}_\mu(x)
\exp\left[{i\sum_{k,a}\tilde{b}_k v^{x,\mu,a}_{k}(\alpha = 0)t_a}\right]\right.\\[2ex]
\left|\vphantom{\sum_{k, a}}
 \tilde{b}_k\in\C,\; x\in\Lambda,\; \mu = 1,\ldots,d
\right\}\,.
\end{multline}
The $k$ index enumerates the different zero-eigenvectors
$v$ of matrix $M$.
For $\tilde{b}_k\in\R$, this is 
a compact manifold. In 
complexified space with $\tilde{b}_k\in\C$, we have non-compact
imaginary directions. Nevertheless, 
the manifold is still critical.
To obtain the generalized Lefschetz thimble
\cite{Witten:2010cx}, we 
start with a compact submanifold
(a cycle) of real dimension $\#(\alpha=0)=n^{(0)}$.
Therefrom its tangent space is spanned. 
The compact submanifold is commonly called
\emph{gauge orbit}. At every point of this cycle,
we use the $\alpha\neq 0$ Takagi vectors to span the
rest of the tangent space. Since the latter 
is invariant under 
zero-modes a point on this
tangent space can be directly written 
in terms of
\begin{multline}
\label{tangent_point}
    U_{\mu}(x) = U^{\mathrm{crit},0}_\mu(x)
    \exp\left[{i\sum_{k,a} b_k c_k v^{x,\mu,a}_{k}t_a}\right], \\
    \text{with}\;\;
    b_k\in\R \; \text{and}\;\;
    c_k = \begin{cases}
    1  & ,\;\alpha_k = 0 \\
    \sqrt{\frac{\sign(\alpha_k)\beta^*}{|\beta |}} &
    ,\;\alpha_k \neq 0\,.
    \end{cases}
\end{multline}
Here the summation index $k$ runs over all eigenvectors. 
The real dimension $n$ 
of the thimble thus splits into zero and non-zero mode directions as
$n=n^{(0)}+n^{(+)}=dVN_g$, 
where $N_g$ is the dimension 
of the Lie algebra. 
This construction can be generalized to 
more complicated Hessians, 
see App.~\ref{GenHessians}.

\begin{figure}[t]
    \centering
    \includegraphics[width=0.4\textwidth,height=0.2\textwidth]{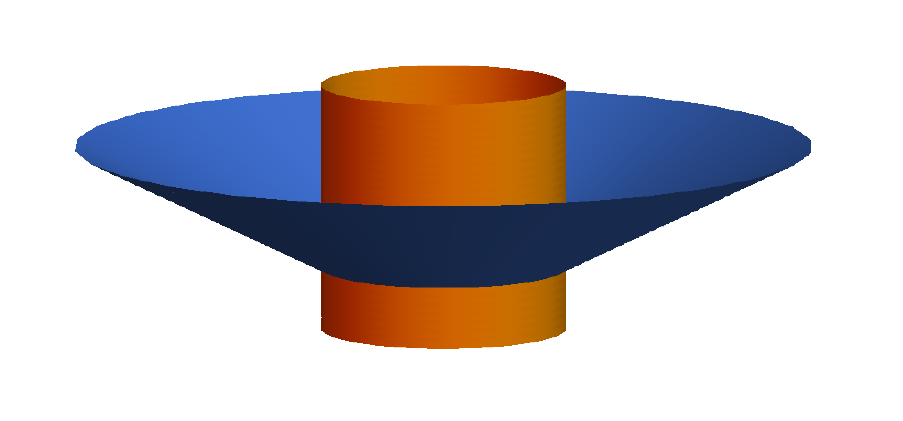}
    \includegraphics[width=0.4\textwidth, height=0.2\textwidth]{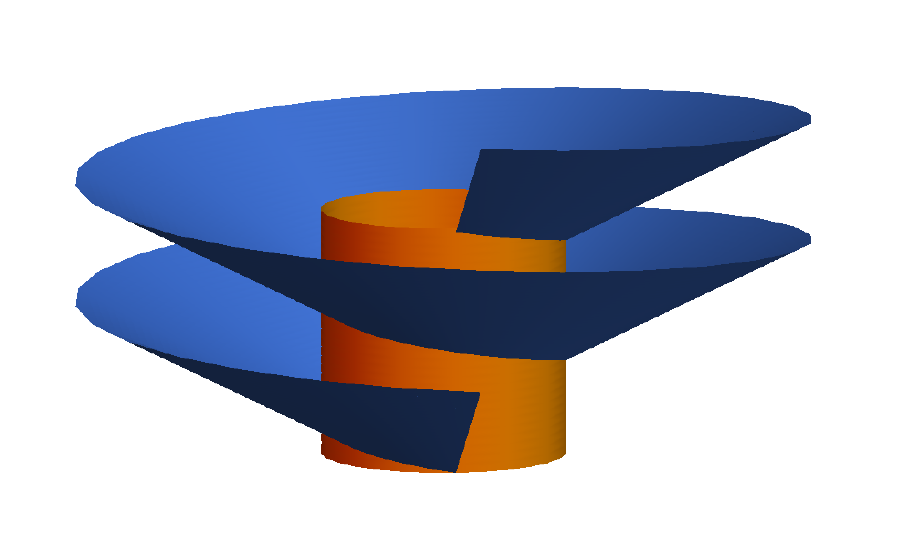}
    \caption{Upper plot: Schematic picture of the generalized Lefschetz thimble, spanned on a compact submanifold as proposed by Witten \cite[3.3]{Witten:2010cx}. Lower plot: The thimble will exhibit infinitely many Riemann surfaces if we choose a noncompact submanifold from the critical cylinder. In both plots the critical cylinder is depicted in orange and the thimble in blue. }
    \label{fig:genThimble}
\end{figure}

If we do not conform to that construction, \textit{e.g.}, tilt the real vectors by a
complex factor, we loose homotopy to the generalized Lefschetz thimble. The 
resulting manifold is non-compact, see Fig.~\ref{fig:genThimble}. 
However, the non-compact directions refer to
zero-modes, \textit{i.e.} they leave the action invariant. This corresponds 
to multiplying the partition sum with an infinite volume factor, which drops out
for all physical observables, when calculating expectation values.
The prerequisite for these observables is, 
that they are independent from the zero-modes, \textit{i.e.} that they are
gauge-invariant observables.

The global minimum among our critical manifolds is
given by $\{P_{\mu\nu}(x)=\mathbbm{1}\;\forall \; x, \mu<\nu\}$.
Since this is a minimum in the non-complexified gauge theory, 
the matrix $M$ is positive semidefinite. As the gauge field is constant,
the complex prefactor is identical to all Takagi-modes.
We are free to choose the same complex prefactor also for the real
zero-modes. Since the eigenvectors of $M$ span the whole
$\R^{n}$ space and since we chose the same complex prefactor for 
all directions, we can make a basis transformation to
the unit basis $\{e_1,\ldots,e_{n}\}$. Moving in one of these 
directions $e_i$ corresponds to a change of a 
single (local) color degree of freedom. 
This basis can thus be used
to construct a local update algorithm on the
generalized main Lefschetz thimble, see Sec.~\ref{sec:AltUpdates}.

As we are interested in a single homotopic covering of our original compact 
integration domain, we need to limit the tangent spaces.
\begin{figure*}[t]
    \centering
    \includegraphics[height=6cm]{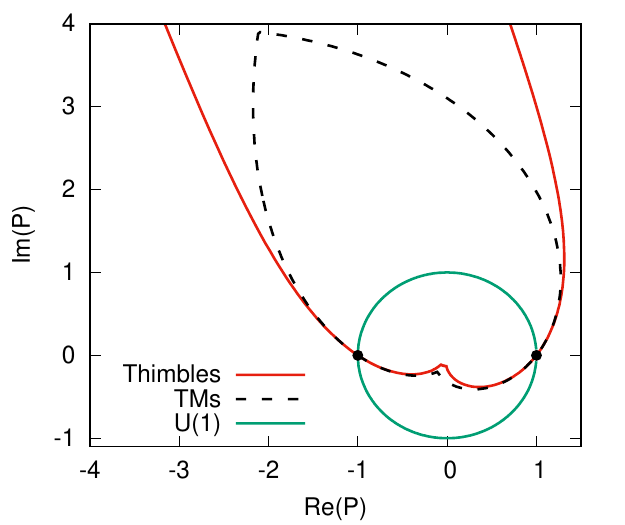}
    \hspace{0.5cm}
    \includegraphics[height=6cm]{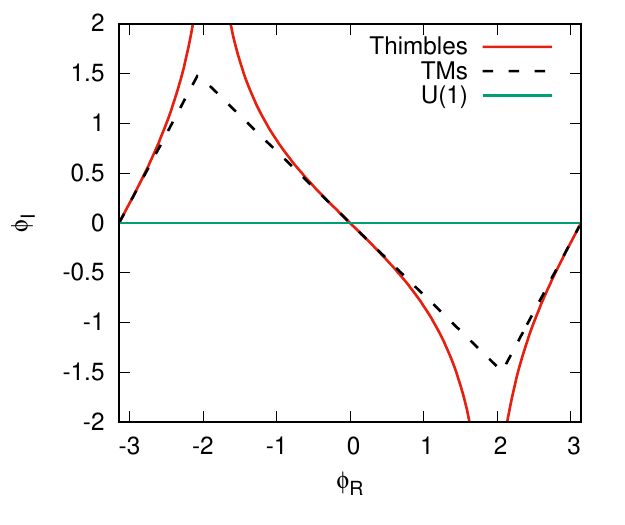}
    \caption{Lefschetz thimbles and 
    tangential manifolds (TMs) bounded by their
    intersection for the one-plaquette model at
    $\beta = 1 + 3i$ in group and angular
    (algebra)
    representation.}
    \label{fig:OnePlaqTangs}
\end{figure*}
It turns out that the construction of a continous  manifold that is homotopic to the original integration domain is the main conceptual difficulty
in this approach.
For the $\UOne$ one-plaquette model, we have only two critical points and tangent 
spaces. They give, when glued together at their intersections, a manifold 
homotopic to $\UOne$, see Fig.~\ref{fig:OnePlaqTangs}.
We introduce boundaries on the main tangential manifold by
identifying intersections with the tangential spaces of the other subleading critical manifolds. We will call these limited tangent spaces, \emph{tangential manifolds (TM)} from now on.
For the full lattice theory we discuss several 
possible choices of boundaries throughout this work.

\subsection{Hierarchy of critical manifolds}
\label{sec:hierarchy}
The choice of critical manifolds introduces a natural hierarchy
which is reflected by the values of the action. Their importance decreases  
according to their weight factor $e^{-S}$ with increasing action. Given our choice of 
critical configurations with diagonal plaquettes from the original integration domain 
($\UN$ or $\SUN$), we can easily express the action in terms of the plaquette eigenvalues.
We find
\begin{eqnarray}
 S = \beta\sum_{x}\sum_{\mu<\nu}\left[1-\frac{1}{N}
\sum_{k=1}^{N}\cos(\phi_{\mu,\nu}^{(k)}(x))\right]\,,  
\end{eqnarray}
where the $\phi_x^{(k)}$ is the angle of the $k$-th eigenvalue of the
plaquette $P_{\mu\nu}(x)$. These critical action values are minima of
the attached thimbles, since the real part of the action naturally
increases if one moves away
from the critical manifolds for $\real(\beta)>0$ in Takagi direction.
This is still true, if one considers the Thimble tangent space
in a region around the critical manifold, limiting it to a TM.
The main critical point ($\phi_x^{(k)} = 0\; \forall x,k$) defines the global minimum of the action in this many TMs scenario.
Consequently with increasing $\real(\beta)$, certain TMs become exponentially suppressed
and we obtain a pronounced hierarchy with the main TM as
leading order.
For purely imaginary values of $\beta$, every thimble contributes equally.

\subsection{Update algorithm on a TM with a Takagi basis}
\label{sec:GenTakagiUpdates}
Next, we discuss possible sampling algorithms that are restricted to a single TM.
Following our strategy to span the tangent space by the Takagi vectors
and \emph{real}
zero-modes, we readily have a parametrization at hand. According to
Eq.~(\ref{tangent_point}) we can express each configuration on the 
tangent space by real coordinates $b_k$, specifying a vector in the Lie-algebra. 
Note that for the zero-modes ($\alpha=0$) we have no complex prefactor. 
Using these coordinates, one can think of  
applying various different update procedures on the tangent space, starting from a 
single random walk Monte Carlo (crude Monte Carlo), over a Hybrid Monte Carlo to a
trained flow-based neural network \cite{Kanwar:2020xzo, Boyda:2020hsi}. 
The important steps are:
\begin{enumerate}
    \item Choose a critical configuration, to specify the tangent space on which to carry out the updates. This configuration may also serve as starting configuration. 
    \item Calculate the real Hessian $M$ and determine its eigenpairs ($\alpha_i$, $v_i$). This has to be done only once 
    for a given critical configuration, independent of the coupling $\beta$.
    \item Propose a new configuration by drawing a set of real coordinates $b_k$. 
    The proposed configuration is than specified according to
    Eq.~(\ref{tangent_point}). Perform an accept/reject step based on the 
    real part of the action difference between the old and new configuration.
    If the proposed configuration is outside the boundaries of the TM, we assign to
    it a zero probability and reject the proposed configuration, recording the old
    configuration like in the Metropolis algorithm.
    \item Finally take the remaining sign problem into account by reweighting with the 
    imaginary part of the action.
\end{enumerate}

\subsection{Leading and subleading thimbles}
\label{sec:algo-subl}
So far, we have restricted the sampling to a single TM attached to a specific critical 
configuration. Ultimately, we are interested in sampling a compact manifold that is homotopic to the 
original integration domain. In the standard thimble decomposition the combination of various Lefschetz 
thimble leads, by definition, to a multi-modal probability distribution. Sampling such distributions by 
a Monte Carlo procedure is difficult. As a solution, a tempered sampling procedure was proposed
\cite{Alexandru:2017oyw, Fukuma:2019wbv}. Another possibility are independent Monte Carlo processes on 
each thimble. However, in this case the relative weights between thimbles need to be known. One possibility
to infer these values is by using prior knowledge of a physical observable for normalization
\cite{DiRenzo:2017igr}.

Here we construct a homotopic manifold by piecewise definition, where we use the TMs as
building blocks. As our construction deviates from the thimble decomposition especially close to the
boundaries, we do not have 
infinite action barriers between the patches. Hence, a sampling
procedure that proposes configurations across boundaries would be in principle possible. However, we found that it 
is
most convenient to sample the main tangent space with a single Monte Carlo chain and take all remaining
patches into account via reweighting. We exemplify this procedure for a system
of two TMs, $\tau_0$, $\tau_1$. 
Calculating expectation values over this extended region requires the relative weight $Z_1/Z_0$. Here
$Z_1$ denotes the partition function corresponding to the subleading 
tangential manifold
and $Z_0$ refers to the corresponding 
quantity on the main tangential manifold. It holds
\begin{align}
 \label{RewObsGen}
\left<\mathcal{O}\right>_{\tau_0\cup\tau_1} &=
\frac{\int_{\tau_0}\dU~\mathcal{O}[U]e^{-S[U]} + \int_{\tau_1}\dU~\mathcal{O}[U]e^{-S[U]}}
{\int_{\tau_0}\dU~e^{-S[U]} + \int_{\tau_1}\dU~e^{-S[U]}} \nonumber \\[1ex]  
&=\frac{\left<\mathcal{O}\right>_{\tau_0} + (Z_1/Z_0)\left<\mathcal{O}\right>_{\tau_1}}
{1+(Z_1/Z_0)}\, .   
\end{align}
Following the method proposed in \cite{Bluecher:2018sgj}, we introduce a mapping
\begin{equation}
    f\colon \tau_0  \longrightarrow\;\; \tau_1\,.
\end{equation}
It maps configurations from one of the two patches to the other. With this mapping we can 
express the ratio $Z_1/Z_0$ as 
\begin{align}
    \frac{Z_1}{Z_0} &= \frac{\int_{\tau_0}\dU e^{-S[f(U)]+S[U]}\det[\mathrm{d}f]e^{-S[U]}}{\int_{\tau_0}\dU e^{-S[U]}} \nonumber \\[1ex]
 &= \left<e^{-S\circ f+S}\det[\mathrm{d}f]\right>_{\tau_0}\,.
\end{align}
It remains to find a suitable $f$. Since we consider only tangent spaces, $f$ is linear
and can be constructed as a basis transformation from the Takagi basis of $\tau_0$ to $\tau_1$.
The Jacobian $\det[\mathrm{d}f]$ is therefore a
constant factor. As the eigenbases of 
the real Hessian $M$ are orthogonal and can be chosen to have determinant one, the 
Jacobian depends consequently only on the the different sets of complex prefactors $c_k$ of the Takagi vectors. These depend in turn only on the sign of the eigenvalues $\alpha_k$. 
Therefore we find for the Jacobian
\begin{equation}
    \det[\mathrm{d}f]
    = \frac{\left(\sqrt{\frac{-\beta^*}{|\beta |}}\right)^{n_{\tau_1}^{(-)}}}
    {\left(\sqrt{\frac{+\beta^*}{|\beta |}}\right)^{n_{\tau_0}^{(+)}-n_{\tau_1}^{(+)}}}\,,
    \label{JacobianFactor}
\end{equation}
where $n^{(+)}$ and $n^{(-)}$ are the number of positive/negative eigenvalues of the real
Hessian $M$ (see Sec.~\ref{sec:lefschetz-and-tang}) at the respective patch. 
Here we assumed that $\tau_0$ is the patch attached to the
main critical manifold having only positive eigenvalues being the 
global minimum.
If the patches $\tau_0$ and $\tau_1$ are of different size, one may introduce an independent
scale parameter to the mapping $f$, which is than also reflected as factor in the Jacobian. 
As the main tangential manifold is usually the largest such a factor is not necessarily needed 
in practice. It may however be used for optimization purposes.

\subsection{Alternative updates on the main tangent space}
\label{sec:AltUpdates}
As outlined in Sec.~\ref{sec:lefschetz-and-tang} 
we can sample the main tangent space
by just setting each complex factor $c_k$ to $\sqrt{{\beta^*}/{|\beta |}}$.
In practice this tilts every link, 
since in this case all
eigenvalues are positive or zero. Having identical complex prefactors, we can perform
a basis transformation of our coordinates to the unit basis. Therein each coordinate 
corresponds to a single gauge degree of freedom. 
This allows for applying a
local heat bath or Metropolis algorithm. The situation discussed here is shown
at the bottom of Fig.~\ref{fig:genThimble}. The unbounded critical manifold in
imaginary direction has to be dealt with.

We limit the plaquette values in accordance with 
the intersection points of the tangent spaces in the one plaquette-model, creating TMs (see Fig.~\ref{fig:OnePlaqTangs}). Similarly as before, we define the region outside the boundary to have zero probability.

But since link variables can still diverge in imaginary direction we have to apply a second limit or just record variables, which are unaffected by the zero modes. In our theory, these are the plaquettes variables. An alternative is to adopt gauge cooling \cite{Seiler:2012wz} to make this problem milder. A severe limitation of this method stems from the fact that only observables invariant under all zero-modes can be measured. Taking pure Yang-Mills theory it is not possible
to measure the Polyakov loop. It is invariant under gauge transformations but not under a global zero mode. The latter is represented by changing all links in the same direction and amount leaving the plaquettes and therefore the action invariant.

\section{Application to a 2-dimensional \texorpdfstring{$\text{U}(1)$}{U(1)}-gauge theory \label{sec:application}}

In this section we apply the above outlined method to
two-dimensional pure $\UOne$ gauge theory with periodic boundary conditions. 
Recently, complementary 
studies on this theory 
with a sign problem have appeared in the literature. 
In \cite{Hirasawa:2020bnl} the complex Langevin method is employed.
The complex action here is caused by a
non-zero vacuum angle.
On the other hand in \cite{Kashiwa:2020brj} 
the theory with a complex gauge coupling is investigated 
by means of the path-optimization method.

\subsection{The effective degrees of freedom}
\label{sec:toron-formulation}
\begin{figure}[t]
\centering
    \includegraphics{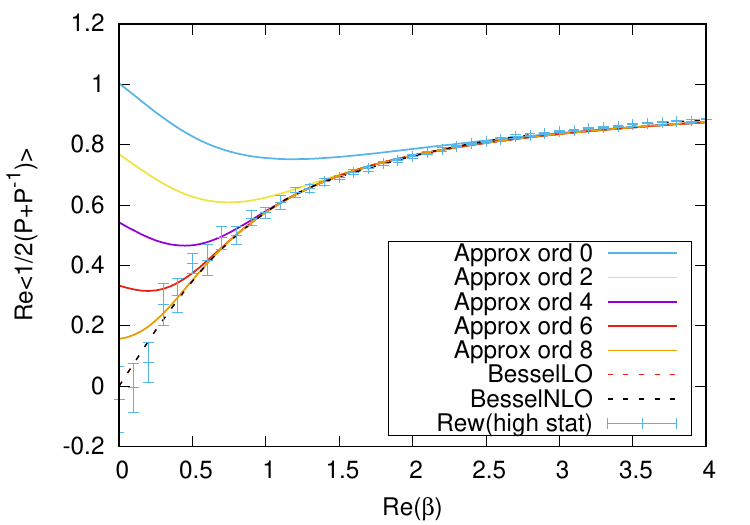}
    \caption{Thimble hierarchy depending on $\real(\beta)$ in the approximation
    on a $4 \times 4$ lattice at constant $\imag(\beta) = 1$.}
    \label{FigApproxHierarchy}
\end{figure}

We formulate the theory in terms of its effective degrees of freedom.\\
Eq.~(\ref{PerCon}) can be verified 
easily
by noting that every link appears
twice in all plaquettes.
Hence the links cancel each other
when being multiplied altogether. 
Consequently, one 
plaquette can be expressed in terms of all others. 
The action of the
two-dimendional theory is rewritten as follows
\begin{eqnarray}
\label{ToronAction}
S & = & -\frac{\beta}{2}\left[\sum_{(x,t)\neq (0,0)}
\left(P_{1,2}(x,t) + P_{1,2}^{-1}(x,t)\right)\right.\\[1ex]
\; & + & \left. \left(
\prod_{(x,t)\neq(0,0)} P_{1,2}^{-1}(x,t) + 
\prod_{(x,t)\neq(0,0)} P_{1,2}(x,t) \right)\right]\,,\nonumber
\end{eqnarray}
neglecting constant terms.
The last term is called \emph{toron} 
term. One can reformulate the full theory in terms
of these plaquettes variables having a reduced
partition sum, which gives the same expectation
values for observables that are invariant under
zero modes
\begin{align}
 & Z \; = \; \int
\prod_{(x,t)\neq (0,0)}\mathrm{d}\theta(x,t) \nonumber \\[2ex]
& \; \times \; \exp\left[\beta/2 \sum_{(x,t)\neq (0,0)}\left(e^{i\theta(x,t)} + e^{-i\theta(x,t)}\right)\right]\\[3ex]
& \; \times \;
\exp\left[\beta/2\left(e^{-i\sum_{(x,t)\neq (0,0)} \theta(x,t)} + e^{i\sum_{(x,t)\neq (0,0)} \theta(x,t)}\right)
\right]\,.  \nonumber
\end{align}
For a full derivation see App.~\ref{app:toron}.
The periodic boundary conditions are 
represented by the \emph{toron} term
we placed at position $(0,0)$. 
Replacing this term
by an independent plaquette term is equivalent
to employing open boundary conditions.  This shall be used
as an approximation to the theory in the following. With open boundary conditions the
integral factorizes in plaquettes variables yielding
\begin{align}
Z = \left[\int_{\UOne}\dP e^{\beta/2(P+P^{-1})}\right]^V = \left[I_0(\beta)\right]^V\,.
\end{align}
The plaquette expectation value is then
\begin{equation}
<\frac{1}{2}(P+P^{-1})>  = 
 \frac{I_1(\beta)}{I_0(\beta)}\,,
\end{equation}
which is exactly the the same as for the one-plaquette model. There is no volume dependence. Since the difference between periodic and open boundary conditions
vanishes in the infinite volume limit, this is the expected
value.\\
We construct the approximation scheme by successively including TMs as integration domains. Thereto we split the integral according to the critical points $P=\pm1$
of the one plaquette model and get
\begin{eqnarray}
Z & = & \left[\int_{\tau_0}\dP e^{\beta/2(P+P^{-1})} + \int_{\tau_1}\dP e^{\beta/2(P+P^{-1})}\right]^V
\nonumber \\
& =: & \left[Z_0 + Z_1\right]^V
 = \sum_{k=0}^V
\begin{pmatrix}
V \\
k
\end{pmatrix}
Z_0^{V-k}Z_1^k\,.
\label{eq:U1approx}
\end{eqnarray}
We can map this to the lattice by $k$ denoting the number
of plaquettes being $-1$ at the critical configuration.
$\begin{pmatrix}
V \\
k
\end{pmatrix}$
is the number of such combinations on the lattice.
Eq.~(\ref{eq:U1approx}) allows to calculate approximate values for the comparison 
with numerical 
simulations (see Fig.~\ref{FigApproxHierarchy}).

Complementary, there exists a formal solution for the for the full lattice theory with periodic boundary conditions involving no approximations. We can write the partition sum
as
\begin{equation}
Z = \int\dU \exp\left({-S[U]}\right) = 
\sum_{n=-\infty}^{+\infty}\left[I_n(\beta)\right]^V\,,
\label{MBesselExp}
\end{equation}
being a series in modified Bessel functions $I_n(\beta)$, where $V$ is the
number of plaquettes \cite{Balian:1974ts, Migdal:1975zg}.
The leading order of this series corresponds to our
approximation. The following orders take finite volume
effects into account and yield the exact result provided that the series converges for the given value of $\beta$.

\begin{figure}[t]
\includegraphics[height=5cm]{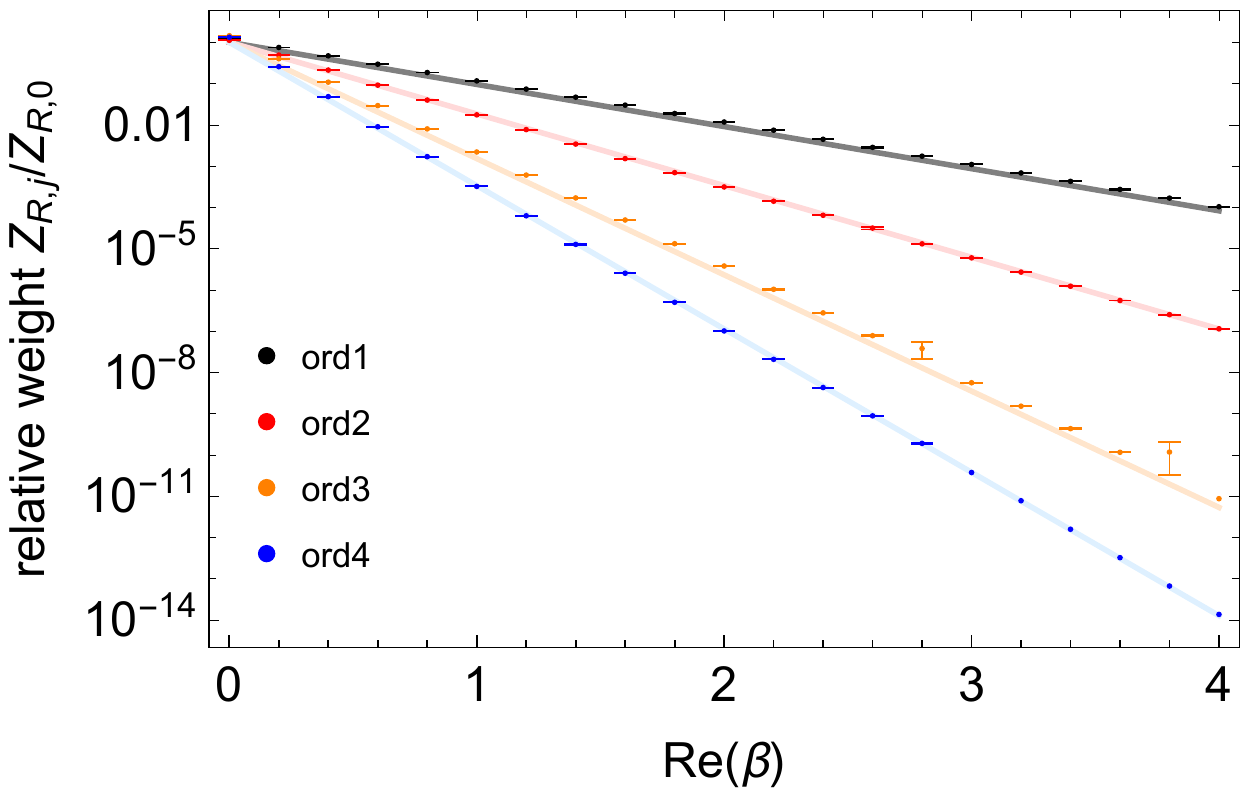}
\caption{Comparison of the ratio of the partition functions 
$Z_{R, j} / Z_{R, 0} = \langle \exp(-(S_{R, j}- S_{R, 0}))\rangle_0$ i.e. the relative weights between the lowest four subleading and the leading order TMs (basis configurations).
The index $j$ labels the subleading orders and the subscript $0$ indicates that the expectation value is calculated on the configurations measured on the leading order tangential surface, as in \cite{Bluecher:2018sgj}. 
Here, the lattice is of size $4 \times 4$, 
$\imag (\beta) = 1$ and spherical boundaries are used.
The data reflects very well the hierarchy 
described in Sec.~\ref{sec:U1-crit-hierarchy}. As a 
comparison the functions $\exp(-S_R[P^{\mathrm{crit}}])$
are added to the plot in light colors. 
}
\label{fig:takagi-Zratio}
\end{figure}
\subsection{Critical manifolds and their hierarchy}
\label{sec:U1-crit-hierarchy}

For an even number $k$ in Eq.~(\ref{eq:U1approx}) the critical configurations in the approximation and in the original lattice theory coincide. In contrast, for odd $k$, this does not hold due to the periodic boundary conditions. However, it is possible to construct critical manifolds being
(arbitrarily) close to the corresponding 
configurations in the approximation. For odd $k < V/2 - 1$ we set $k$ plaquettes to 
$e^{i(\pi-\epsilon)}$ and the rest to  $e^{i\epsilon}$
such that Eq.~(\ref{eq:eps-top-eq}) is satisfied. 
This leads to the choice
\begin{equation}
    \label{eq:basiceps}
    \epsilon = \frac{\pi}{V-2k},
\end{equation}
which we refer to together with the critical configurations for even $k$
as \emph{basis} configurations from now on. 
Furthermore as discussed in Sec.~\ref{sec:ym-crit-2}, Eq.~(\ref{eq:eps-top-eq}) gives rise to a multiplet of configurations being related to the basis configurations by a symmetry.  To see this, note that these $\epsilon$ configuration have topological charge $\lceil\frac{k}{2}\rceil$. Now, we can change the topological sector by adding/
subtracting $\frac{2\pi}{V - 2k}$ from $\epsilon$. We observe that if
$|\epsilon| < \pi/2$, the Takagi vectors do not change. Consequently we can
apply this transformation directly to measured plaquettes values by
multiplying them with $e^{\pm2\pi/(V-2k)}$, where the sign is chosen 
whether we have a $e^{i(\pi - \epsilon)}$ or $e^{i\epsilon}$ plaquettes.
This transformation does not only apply to odd $k$ but also to even $k$.
By this procedure, we reach $2(V/4 - \lceil\frac{k}{2}\rceil)$ critical manifolds
in different topological sectors for odd $k$ and $2(V/4 - 1 - \frac{k}{2}) + 1$
for even $k$, respectively.\\
Second, we can go from the $k$-th TM
to the ($V\!\!-k$)-th TM by changing the plaquettes from the
critical configuration according to
\begin{equation}
P^{\mathrm{crit}}_{1,2}(x) \;\longrightarrow\; -(P^{\mathrm{crit}}_{1,2}(x))^{-1}\,.
\end{equation}
The corresponding real Hessian $M\rightarrow (-M)$ changes
sign and has the same eigenvectors and zero
modes. Only the non-zero eigenvalues $\alpha\rightarrow(-\alpha)$
change sign.
Consequently and using the fact that eigenvectors are unique
up to a non-zero scalar multiplication we get
the Takagi vectors for the $(V\!\!-k)$-th TM by
multiplying the Takagi vectors from the $k$-th TM
by $\sqrt{-1}=i$.
For $k \neq V/2, V/2\pm 1$, all zero modes do not change the
plaquettes. Writing a plaquette configuration on
the $k$-th tangent space as
$P^{\mathrm{crit}}_{1,2}(x)e^{i\Delta\varphi(x)}$, we can write
the mapping to the configuration on the 
opposite $(V\!\!-k)$-th tangent space as
\begin{equation}
P^{\mathrm{crit}}_{1,2}(x)e^{i\Delta\varphi(x)}\;\rightarrow\;
-(P^{\mathrm{crit}}_{1,2}(x))^{-1}e^{-\Delta\varphi(x)}\,,
\end{equation}
having again a transformation for directly measuring the
plaquette values. Having these, the same procedure to reach the different
topological sectors can be applied to these plaquette values.

An exception is the case $k=V/2$, which has an additional zero mode,
which can be parametrized by
\begin{equation}
\label{eq:Vhalfzeromode}
    P_{12}(x) = e^{i\varphi}\;\text{and}\;P_{12}(x) = e^{i(\pi-\varphi)}\,,
\end{equation}
for $V/2$ plaquettes on either side. This transformation necessarily leaves
the action invariant while changing actual plaquettes values.
The configurations, where all plaquettes are $\pm i$, which would be assigned
to $k = V/2 \pm 1$ in our scheme are included in this critical manifold and are therefore being left
out. This transformation reduces the combinatorial factor to 
$\frac{1}{2}\begin{pmatrix}
V \\
V/2
\end{pmatrix}$.

The critical manifolds form a hierarchy depending
on their associated value of the action
\[S \ 
\begin{cases}
=2k\beta,&\: k\; \text{even},\\[1ex]
\approx \beta\left(2k + \frac{\pi^2}{2(V-2k)^2}\right), & \;
k\;\text{odd}\,.
\end{cases}
\]
for the basis configurations.
Depending on $\real(\beta)$ the critical manifolds differ
in importance for the partition
sum since on thimbles and suitably bounded TMs
the real part of the action
is minimal at the critical manifold. 
Consequently, if $\beta$ is purely
imaginary, every thimble or TM contributes equally.
Otherwise one can obtain an approximate
result by taking only a few thimbles or TMs into account
as the others are exponentially suppressed.
Fig.~\ref{fig:takagi-Zratio} illustrates the hierarchy of the critical manifolds considering subleading TMs up to order $k = 4$ restricted to the basis configurations. The simulation setup used for the shown data is described in detail in Sec.~\ref{sec:sampling-main-TM}
and \ref{sec:Subtang}.

\begin{figure}
\includegraphics[width=8cm]{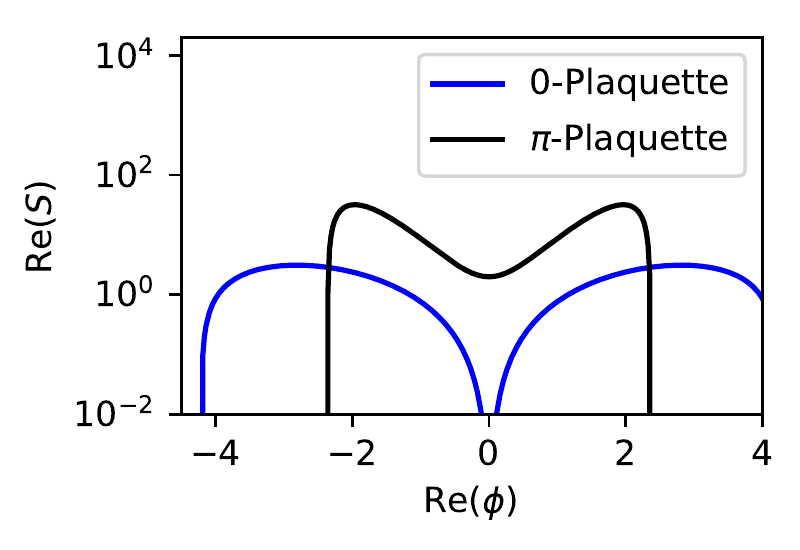}
\includegraphics[width=8cm]{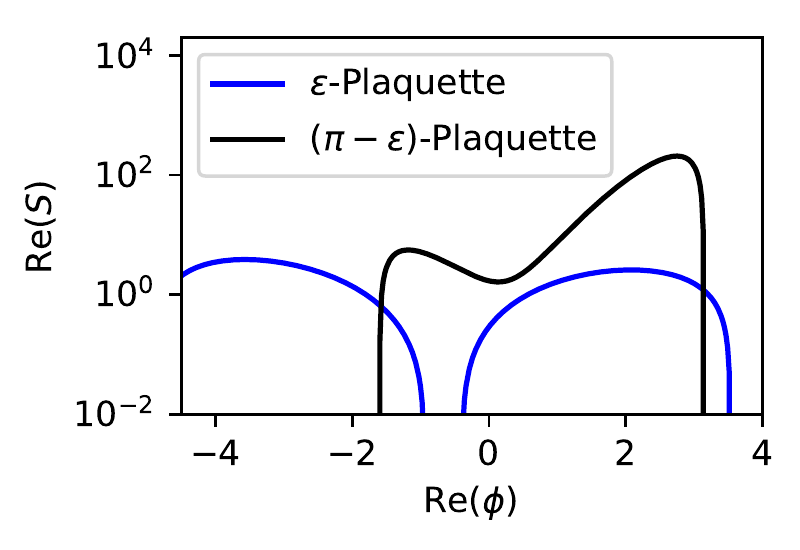}
\caption{Real part of the action of a single plaquette on the tangent space
of the thimble. At the top, we look at the situation $\phi = 0$, i.e.
$P=1$ and $P=-1$ like in the one plaquette model. In the plot below, we applied
a shift $\epsilon = \pi/4$ to see the behaviour of shifted plaquettes, which we encounter
in critical configurations with non-zero topological charge.}
\label{fig:EpsActionPlot}
\end{figure}

\begin{figure*}[t]
    \centering
    \includegraphics[height=6cm]{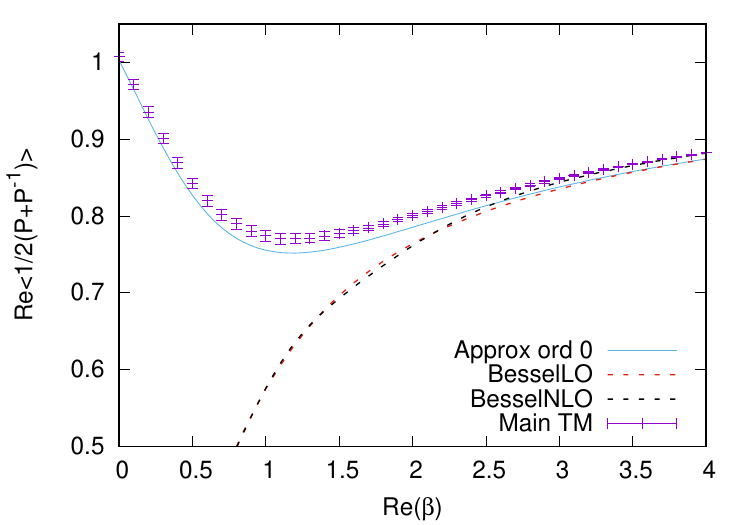}
    \includegraphics[height=6cm]{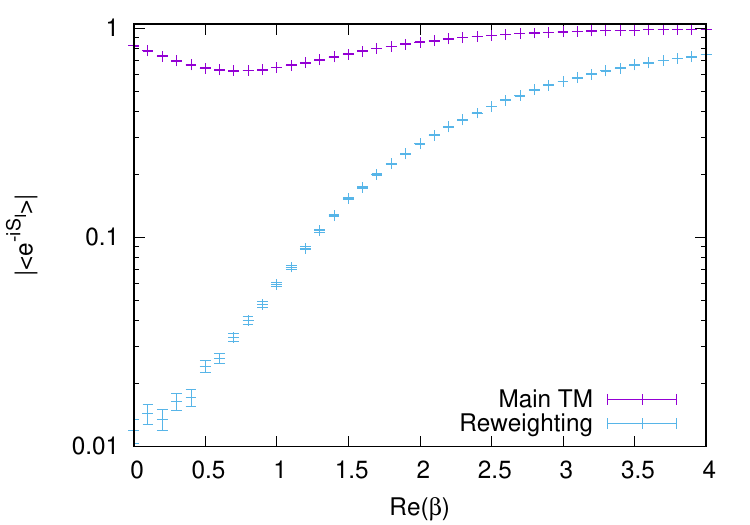}
    \caption{Result of the simulation on the main TM with real
    plaquette boundaries for constant $\imag(\beta)=1$ on a $4 \times 4$ lattice.
    Note the fact that for high $\real(\beta)$, the result approximates
    the BesselNLO result. This indicates that we
    simulate the actual lattice theory and not by accident
    a cartesian product of the one-plaquette model. On the right is the average
    sign compared with standard reweighting.}
    \label{FigMTbetaI1}
\end{figure*}

\subsection{Ensuring homotopy}
\label{sec:ensuring-homotopy}
To ensure global homotopy, we need to make sure, that the TMs
form a patchwork covering $\left[\UOne\right]^{2V}$. Therefore, we have
to create boundaries, which match each other exactly. Strictly speaking,
this is not generally possible in higher dimensions, since there is no theorem
that tells us, that these tangent spaces have to intersect in this
manner as it is the case for thimbles. But we can at least get close to something alike
minimizing the systematic error introduced by homotopy violations
as much as possible.\\
Our approach comes with thinking in effective degrees of freedom being plaquettes variables with 
a toron term
as shown
in Sec.~\ref{sec:toron-formulation}. Looking at the different (subleading) tangent spaces, we have applied several
schemes, each based on different criteria:

 \begin{enumerate}
    \item Real plaquette boundaries based on the one plaquette model:
    These are shown in Fig.~\ref{fig:OnePlaqTangs}. 
    The plaquette variables take values in a region bounded by
    the intersection points of the two tangents in the one-plaquette model.
    In the full lattice theory, we still find these intersection points
    for the transitions in configurations where $k$ is even.
    We therefore limit the real parts of the plaquettes by these intersections.
    For $\epsilon \neq 0$ configurations the transition looks different
    and we take it into account by a shift of the boundaries.
    Having transition points does not mean that we get full homotopy.
    We would need to identify intersections with real dimension $2V-1$ which do not
    necessarily exist.
    \item Imaginary plaquettes bounds: The limit can also be applied to the imaginary part of the plaquettes
    preventing them from drifting to far off in
    imaginary direction.
    This allows a larger space to be explored than for real
    plaquettes bounds. Moreover these boundaries are far easier to implement  since
    all real critical manifolds are part of the original
    group and we only need one value to specify these
    boundaries.\\
    However, this doesn't care so much for homotopy like
    the real boundaries, since the plaquettes in 
    the full lattice theory only approximately lie on 
    the same tangent spaces found in the one plaquette model.
    Consequently overlapping TMs are not excluded in the case
    of imaginary plaquettes boundaries, while the real plaquette
    boundaries still guarantee that there is no Overlapping.
    \item Action boundaries: On a Lefschetz thimble, there exists a coordinate
    system with the critical point at the center, where $S_R$ is a simple
    rising quadratic function \cite[Lemma 2.2]{milnor:morse1969}. Since the TM
    is close to it in the vicinity of the critical manifold, we observe the
    similar rising behaviour up to some distance illustrated
    for the local action in Fig.~\ref{fig:EpsActionPlot}. We limit the plaquettes 
    variables by the local maxima of $S_R$. Since configurations are
    exponentially suppressed by it and the fact that this distance goes
    beyond the intersection points mentioned before allows a larger part
    of configuration space to be explored, while the systematic error stays
    small.
    \item Spherical boundaries: This is the most conservative
    choice being independent from the
    coordinate system. We observe that wandering along
    the Takagi vector with the highest eigenvalue $\alpha = 8$ (this is true
    for all even lattice sizes including and greater than $2 \times 2$)
    for the main TM turns it into 
    the ultimate subleading TM (whose critical
    configuration has all plaquettes equal to $-1$) intersecting with its
    counterpart ($\alpha = -8$) from there. This can be
    understood in terms of the effective d.o.f. as
    a diagonal connection in a hypercube. The intersection
    marks a corner in the cube containing the main TM. We now choose the radius of
    the inner sphere of the cube to make sure, 
    we do not intersect with other TMs, where we
    can assign a radius in the same way. We get
    \begin{equation}
    r_{\mathrm{max}} = \frac{\sqrt{2V}\pi}{4\sqrt{V-1}}\real\sqrt{\frac{\beta}{|\beta|}}\,,
    \end{equation}
    as the radius of the effective sphere (the distance from
    the critical manifold is calculated in terms of the 
    non-zero eigenmodes). \\
    A disadvantage is that here the curse of
    dimensionality hits directly into the calculations, since
    with rising dimension, the volume of the inner sphere becomes
    negligible in comparison to the cube and we explore only small portions of configuration space.
    One can counter that by scaling the radii paying the price
    of overlapping TMs introducing another systematic 
    error.
    \item $\mathrm{Im(S)}$ boundaries: The idea here is that
    the TMs go into the other TMs by their intersections.
    So the imaginary part of the action of one TM has to change continuously to the one of the other (On thimbles
    $\imag S$ is constant and changes abruptly at their 
    singular intersection, which in our case is at infinity.). Having a pronounced hierarchy allows us to 
    set the boundaries using this feature by e.g. allowing
    one tangent space only to vary in an interval of $\imag S$ and the other one on a subsequent interval.\\
    Problematic is the fact that we have intersections 
    of one TM with multiple TMs in different orders. So here
    we possibly limit the explored space too much.
\end{enumerate}

All in all, we have found a combination of real plaquettes
boudaries with $\epsilon$ shifts and action boundaries most
promising. To that end we calculate the former and correct
them, if they extend futher than the action boundaries, which
is especially important for TMs with high $\epsilon$. Otherwise we can fall into the regions with negative action,
where the the TM is far away from the thimble, see Fig.~\ref{fig:EpsActionPlot}.

\subsection{Algorithm for sampling on the main TM}
\label{sec:sampling-main-TM}
In the following we explain the sampling method to generate configurations on the 
main tangential surface. 

\begin{itemize}
    \item[(1)] Diagonalize the Hessian at the main critical point where all plaquettes are $+1$.
    \item[(2)] Construct the parametrization of the main TM surface. 
               The eigenvectors from (1) with non-zero eigenvalues 
               correspond to thimble directions. Tilt those as described in Eq.~(\ref{Takagi}) above to obtain the Takagi basis $\{\xi_i \ i = 1, \hdots, 2V\}$. 
    \item[(3)] Define boundaries. Before running the simulation 
               specify the configuration space to be sampled by 
               choosing suitable boundaries of the main TM. For possible choices see Sec.~\ref{sec:ensuring-homotopy}. 
    \item[(4)] Run the Monte Carlo simulation. A vector on the leading TM in the Takagi basis is given by $\xi = \phi_i \xi_i$ with $\phi_i \in \mathbb{R}$. 
    Beginning with a cold start at the critical manifold, i.e.~$\phi_i = 0 
    \ \forall \ i = 1, \hdots, 2V$ sample the $\phi_i$ via 
    the Metropolis algorithm using proposals $\Delta \phi_i \sim \mathcal{N}(0, \sigma)$. Here, a sweep is defined by applying a Metropolis accept-reject
    step for every direction $i$. The Metropolis updates are constructed 
    such that the following conditions are satisfied
    \begin{itemize}
    \item[(i)] Configurations with $S_R < 0$ are being rejected.
    \item[(ii)] A proposed configuration outside of the specified boundaries is being rejected.
    \end{itemize}
\end{itemize}

The measured configurations are stored to be used for the reweighting on the subleading TMs. This part of the algorithm is described in detail in Sec.~\ref{sec:Subtang}. 

Like already mentioned in Sec.~\ref{sec:AltUpdates}, there is also a local update algorithm (see App. \ref{AltUpdatesU1}).

\subsection{Numerical results on the main TM}

Looking at the approximation in Fig.~\ref{FigApproxHierarchy}, we expect the main tangent results to roughly follow the zero order
approximation. For high $\beta_R$, the other tangent spaces are
exponentially suppressed allowing for convergence to the full result.
Noticing that the full result for this range
is slightly above the one plaquette model due to finite volume
effects, we hope to see the same from the simulation, which
is the case, see Fig.~\ref{FigMTbetaI1}. This proves that
we do not accidentially just simulate the one-plaquette model
by our procedure. 

Another important point, is that the sign problem should be reduced
in comparison with standard reweighting, which is also the case (see second plot in Fig.~\ref{FigMTbetaI1}). 

So, we can expect for high enough $\beta_R$ (e.g. here larger than $3$) correct results for simulations at complex beta with a lesser sign
problem. To extend this range, we need to take the subleading
TMs into account, see Sec.~\ref{sec:Subtang}.

\begin{figure}[t]
    \centering
    \includegraphics{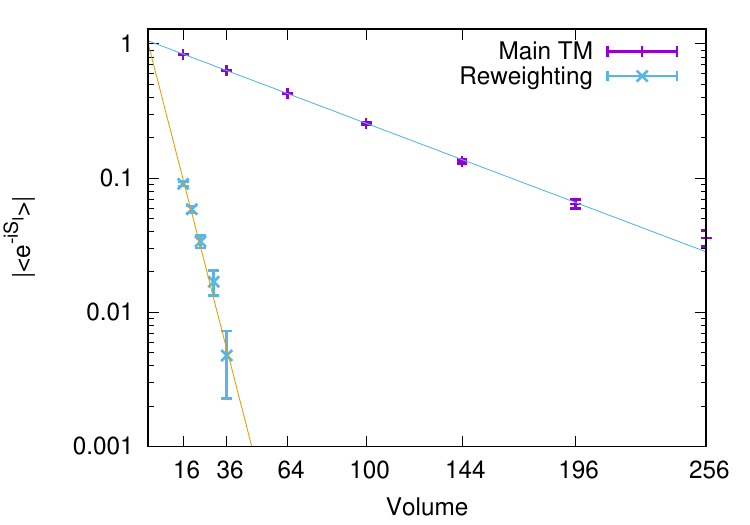}
    \caption{The average sign plotted for increasing volume
    at constant $\beta = 2+1.4i$.}
    \label{fig:my_label}
\end{figure}

For standard phase reweighting, the average sign should exponentially decrease with
increasing space-time volume. Since our simulation works similarly
just on a tilted space, we see the same happening here. The difference
is that our average sign is higher than the one for
standard reweighting and the slope is less steep, see Fig.~\ref{fig:my_label}. Therefore, higher lattice volumes are more easily accessible in our approach. Indeed, we needed to increase 
statistics for the reweighting simulations, while the number of samples for the different volumes in the TM simulations remained the same.

\begin{figure}[t]
\includegraphics[height=6cm]{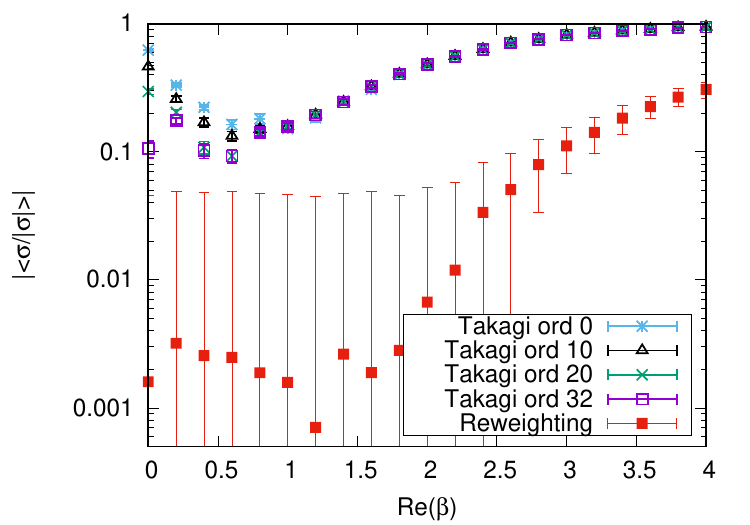}
\caption{Absolute values of the expectation values of the
phase of the reweighting factor of Eq.~(\ref{RewObsTak})
depending on the number of included tangent spaces
and $\beta$ in comparison with a standard reweighting 
simulation.
}
\label{fig:takagi-rew-sign}
\end{figure}

\begin{figure*}[t]
\includegraphics[height=6cm]{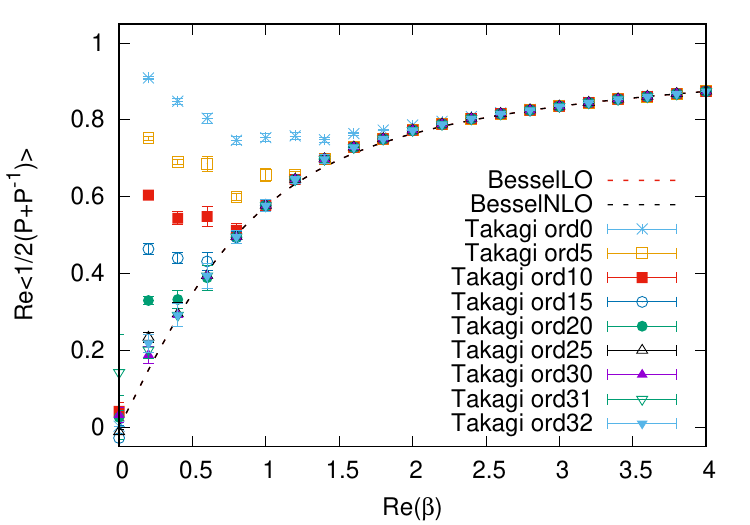}
\hspace{0.5cm}
\includegraphics[height=6cm]{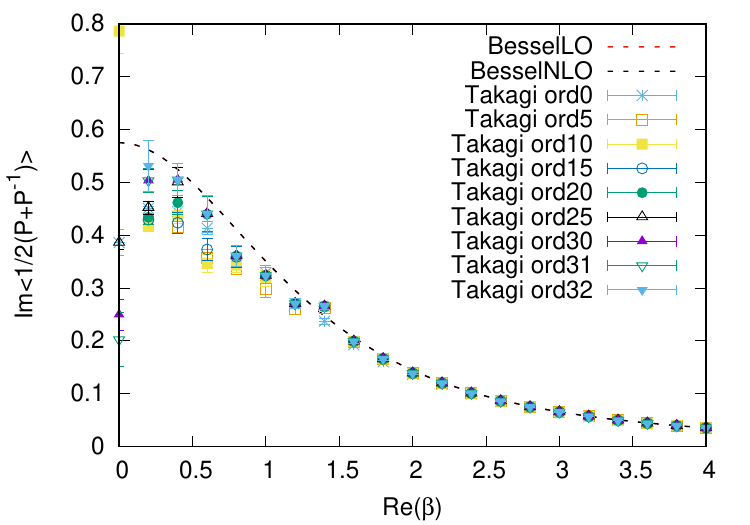}
\caption{Takagi simulation incorporating
different orders of reweighted tangent spaces on an $8 \times 8$
lattice
at constant $\imag(\beta)= 1$.
The dashed line indicates the Bessel result, see Eq. (\ref{MBesselExp}). Each order $k$ contains
for practical reasons the $V-k$ order and all topological
TMs associated with these. 
Shown are the real part and imaginary part of the plaquette expectation value.}
\label{fig:takagi-rew-results}
\end{figure*}

\subsection{Reweighting of subleading TMs\label{sec:Subtang}}
For the reweighting, the observable from Eq.~(\ref{RewObsGen}) becomes
\begin{align}
    \label{RewObsTak}
    & <\mathcal{O}> \;= \\[1ex] \nonumber 
    & \quad\frac{< e^{-iS_I}\mathcal{O} + \sum_{k=1}^{n}
    m_k \det[\mathrm{d}f_k]e^{S_R-S\circ f_{k}}
    \mathcal{O}\circ f_{k} >_{\tau_0}}
    {< e^{-iS_I} + \sum_{k=1}^{n}
    m_k \det[\mathrm{d}f_k]e^{S_R-S\circ f_{k}} >_{\tau_0}}\,,
\end{align}
where $n$ denotes the number of subleading tangent space orders, one
wants to incorporate, $m_k$ are the multiplicities and
$f_k$ are linear transformations projecting
the main TM $\tau_0$ onto the subleading tangent space
$\tau_k$. This is done by aligning the Takagi basis using
the real vectors $v_j$ to calculate a transformation matrix
\begin{equation}
    \gamma^{(k)}_{ij} = (v^{(k)}_j)^Tv^{(0)}_i\,,
\end{equation}
which we apply to the subleading Takagi basis
\begin{equation}
    z^{(k)}_i = \sum_{j} \gamma^{(k)}_{ij} c^{(k)}_j 
v^{(k)}_j\,,
\end{equation}
where the $c^{(k)}_j$ are the complex prefactors from
Eq.~(\ref{Takagi}). The $z^{(k)}_i$ are now
our new aligned basis vectors for $\tau_k$.
Therefore we can directly project from the leading to
subleading tangent spaces. As already mentioned in Sec.~\ref{sec:algo-subl},
depending on the other boundaries of the subleading tangent space, one can
apply scaling factors to the variables.
We observed, that for $\beta_R, \beta_I \geq 0$, the main TM is always
larger (or equally large) as the other subleading TMs.
Instead of rescaling, we use an indicator function
$\chi_{\tau_k}(U)$ for the boundaries:
If the projected space $f_k(\tau_0)=\tilde{\tau}_k \supseteq \tau_k$,
then we have
\begin{align}
\int_{\tau_k}\mathrm{d}U g(U)=
\int_{\tilde{\tau}_k}\mathrm{d}U g(U)\chi_{\tau_k}(U)
\end{align}
with $g(U)$ an arbitrary function on the subspace.
So, we simply set the integrand to zero in the reweighting
process, if the projected configuration is out of bounds. 

We use several symmetries to incorporate the
different topological sectors for each order as well
as a mapping from the $k$-th to the $(V\!\!-k)$-th TM
already discussed in Sec.~\ref{sec:U1-crit-hierarchy}.
In the end, we have to diagonalize only $V/2$ Hessians to get
every contribution. For practical reasons, we subsumed
the $V/2$ TM under order $V/2 - 1$ and $V/2$ using
a turn by $\varphi=\pi$ of the additional zero mode 
(\ref{eq:Vhalfzeromode}). During the reweighting process,
we need to check for every contribution, if it is in its predefined
boundaries, since they also differ over topological sectors (see 
Sec.~\ref{sec:ensuring-homotopy}). 

We applied the procedure on an $8 \times 8$ lattice at
constant $\imag(\beta) = 1$ and compared 
different orders of reweighting with the BesselNLO result
(see end of Sec.~\ref{sec:toron-formulation}). The results
are shown in Fig.~\ref{fig:takagi-rew-results} as well as the average
sign in Fig.~\ref{fig:takagi-rew-sign} depending on how many
orders of TMs one takes into account. The boundaries chosen are
real plaquette boundaries based on the one plaquette model, since
they prevent overlapping of the TMs, while allowing a large space to
be explored.

\section{Discussion\label{sec:discussion}}
As we have stated already in the introduction, the particular choice of our deformation has two important advantages:
\begin{itemize}
\item[($A1$)] Due to the flatness of the patches a parametrization in terms of real coordinates and
basis vectors is easily constructed. It is thus straight forward to realize a sampling
procedure on a particular patch. In contrast to the generalize Lefschetz thimble approach,
there is no need to solve a flow equation to propose a new configuration nor to evaluate a
Jacobian at each sampling point. 
\item[($A2$)] Since our coordinate systems originate from a critical configuration
on each patch, which is the configuration that receives the largest weight on that patch, 
a reweighting from one patch to another is possible without any severe overlap problem. 
\end{itemize}
In the case of a pronounced hierarchy among critical configurations, we have demonstrated 
in the case of the 2d U(1) theory, that a well controlled approximation scheme emerges if
successive contributions from suppressed patches are taken into account. 

On the other hand there are two -- as we believe less severe -- disadvantages: 
\begin{itemize}
\item[($D1$)] Sampling on the flat tangential manifold rather than on the curved thimble, does not
erase the sign problem completely. It would be interesting to analyze whether the optimal
deformation of the original manifold, which minimizes the combined sign problem of the action
on the integration domain and the one introduced by the Jacobian, is closer to the thimble
decomposition or our decomposition from flat patches. We leave this for future investigations.
We have demonstrated that the resulting sign problem in the case of the 2d U(1) gauge theory
is, although sill exponential in volume, very mild compared to the standard reweighting
procedure. 
\item[($D2$)] In order to construct our deformation, we need to know all relevant critical
configurations. Depending on the space-time dimension, volume and gauge group, this can be a
very large amount of critical configurations. 
\end{itemize}
It is one of the main results of this work that we have identified a large number
of critical configurations, which are characterized by specific distributions of plaquettes
across on the lattice. 
In particular, we have demonstrated that we can chose these critical configuration from the
maximal torus of the original gauge group. We have shown further that there exists a large number
of degenerate critical configurations and patches due to lattice symmetries. For the reweighting
procedure we thus need to take only one of these patches into account if appropriate 
combinatorial multiplicity factors are used. The results look promising and systematic
errors by non-homotopy vanish with growing lattice size.

We want to emphasize here that the choice of a complex coupling $\beta$ is not at all a
pathological choice. In the limit of a purely imaginary coupling, the kernel of the discussed
partition sum can be seen as the real time propagator of the theory. For the evaluation of
real time correlation functions in a thermal bath, the Schwinger-Keldysh formalism is usually
applied. Our sampling strategy might also be applied to the Schwinger-Keldysh contour, even
though in this case an additional sign problem arises from the edges of the contour. 

Moreover, the critical configurations we have identified here are not only critical in the case
of the Yang Mills action with complex coupling $\beta$. The same configurations remain
critical when we introduce fermionic matter fields with a chemical potential $\mu$. In this
case the effective action might be written as 
\begin{equation}
    S_{\mathrm{eff}}(\mu;U)=\beta S_G(U) - \trace \ln D(\mu;U)\,.
\end{equation}
Hence, the action gradient is the sum of a contribution from the gauge 
($S_G(U)$) and the fermionic part ($S_F(\mu;U)$) of the action. That the action gradient of the 
gauge part vanishes at our critical configurations has been discussed in detail above. The fermionic contribution to the gradient is given as 
\begin{align}
      \partial_{x, \nu, a} S_F= &\,
    i \,\trace \Biggl[ D^{-1}\Bigl (e^{\mu\delta_{\nu,0}} U_\nu(x) \nonumber \\[1ex]
    &\hspace{1.5cm}-e^{-\mu\delta_{\nu,0}} U^\dagger_\nu(x+\hat \nu) \Bigr)t_a\Biggr] \,.
\end{align}
For those critical points that are not only chosen from the maximal torus of the gauge group 
but are also constructed from center elements of the original gauge group,
the fermionic contribution vanishes as well. As all link variables are
proportional to the unit matrix, the (none sparse) inverse of the fermion matrix $D^{-1}$ 
contains $N\times N$ diagonal blocks which are also proportional to the unit matrix. We conclude
that the matrix multiplying the generator $t_a$ is proportional to the unit matrix and as such 
the whole expression vanishes. 
We thus hope that our strategy might also prove useful in this case, \textit{i.e.} the field
theoretical description of dense matter, including QCD at net-baryon number
density.
\section{Conclusion \label{sec:conclusion}}
We have put forward here a novel nonperturbative lattice approach for Yang Mills theories with a complex 
gauge coupling $\beta$. The approach is based on a deformation of the original
integration domain of the theory into complex space. Guided by the Lefschetz thimble
decomposition of the partition sum we have chosen a new integration domain. This is constructed
piecewise from patches of tangential manifolds to the 
relevant Lefschetz thimbles. 

For the numerical implementation we have chosen to set up a Monte
Carlo procedure on the main tangent space only. All further contributions from other
patches are taken into account by reweighting. We have tested our approach by applying it to the case of 2d U(1) gauge theory. Here, it far out-performs our benchmark simulation with standard reweighting, see Fig.~\ref{fig:U1result}. 

Based on this observation we plan to apply our approach to general $\UN$ and $\SUN$
gauge groups in 4d. While our approach has shown to feature an exponentially better performance than standard reweighting, it remains to be shown that the sign
problem stays numerically manageable also in these cases. We hope to address these questions 
in a forthcoming  publication. Moreover, we envisage simulations of fermionic matter fields at finite chemical potential as well as real-time lattice theories with expectation
values along the Schwinger-Keldysh contour.

\section*{Acknowledgements}
The authors thank Andrei Alexandru, Benjamin J\"ager, Alexander
Lindemeier and Ion-Olimpiu Stamatescu for discussions.

C. Schmidt and F. Ziesché acknowledge support by Deutsche Forschungsgemeinschaft (DFG,
German Research Foundation) through the Collaborative Research Centre CRC-TR 211
’Strong-interaction matter under extreme conditions’ project number 315477589 and from the
European Union’s Horizon 2020 research and innovation program under the 
Marie Skłodowska-Curie grant agreement No H2020-MSCAITN-2018-813942 (EuroPLEx). 
This work is further supported by the
ExteMe Matter Institute EMMI, the Bundesministerium für Bildung und Forschung (BMBF, German 
Federal Ministry of Education and Research) under grant 05P18VHFCA and by the DFG through the 
Collaborative Research Centre CRC 1225 (ISOQUANT) as well as by DFG under Germany’s Excellence
Strategy EXC-2181/1-390900948 (the Heidelberg Excellence Cluster STRUCTURES). M. Scherzer
acknowledges support from DFG under grant STA 283/16-2. 
F.~P.~G.~Ziegler acknowledges support by Heidelberg University where a part of this work was carried out. 

\appendix

\section{General complex Hessians}
\label{GenHessians}
If we can not write the Hessian as $H=\beta M$, with a real matrix $M$ as in
Sec.~\ref{sec:lefschetz-and-tang}, we consider real and imaginary parts of the Takagi
Eq.~(\ref{takagi}) separately (see \textit{e.g.} \cite{Alexandru:2015xva})
\begin{multline}
(H^R - iH^I)(v^R - iv^I) = \lambda (v^R + iv^I)\\[1ex] \Leftrightarrow
\begin{pmatrix}
H^R & -H^I \\
-H^I & -H^R
\end{pmatrix}
\begin{pmatrix}
v^R \\
v^I
\end{pmatrix}
= \lambda
\begin{pmatrix}
v^R \\
v^I
\end{pmatrix}\;.
\end{multline}
The matrix is by definition symmetric and we have only real
eigenvalues, positive $\lambda$ for Takagi and negative $\lambda$
for Anti-Takagi vectors. The zero modes span again the whole
critical manifold. Since for compact gauge groups, we are only
interested in the real subspace, which are naturally compact.
Therefore, we impose $v^I=0$ and get
\begin{equation}
    H^Rv^R = 0 \;\;\text{and}\;\; H^Iv^R = 0\;.
\end{equation}
This implies that the critical manifold is spanned by the mutual 
zero modes of the real and imaginary part of the Hessian. 
The strategy therefore is to calculate the zero modes either parts and test, if these are also zero modes of the other part. The
number of real zero modes and of the Takagis have to add up to the overall dimension.
\section{The toron formulation}
\label{app:toron}
We reformulate the theory as mentioned in Eq. 
(\ref{ToronAction}). Then we gauge all time-like directions
into one link. This reduces the degrees of freedom by 
$N_{x}(N_{t} - 1)$.

Next, we define new lattice variables $\theta$ and
express the links (in their algebra representation $\phi_{\mu}(x,t)$) in terms of these.
Thereto we write direction vectors 
($\vec\varphi(x,t), \vec{\varphi}_{2}(x), \vec{\varphi}_{1}(0)$)
in link space indicating
how the new variables change the links in their respective direction. This yields the 
following parametrization
\begin{align}
\label{LinkParam}
    \phi_{\mu}(y,t) & =
    \sum_{(x,\tau)\neq(0,0)}\theta(x,\tau)(\vec{\varphi}(x,\tau))_{\phi_{\mu}(y,t)}  \\[2ex]
    & + \sum_{x}\theta_{2}(x)(\vec{\varphi}_{2}(x))_{\phi_{\mu}(y,t)}
      + \theta_{1}(0)(\vec{\varphi}_{1}(0))_{\phi_{\mu}(y,t)} \nonumber
\end{align}
for each remaining link. The $\theta(x,t)$ shall denote plaquette variables,
while the $\theta_{2}(x)$ and $\theta_{1}(0)$ denote zero modes
at space slices or the zero time slice. We will integrate them out later.
All space-like links
 can be replaced by plaquette variables
\begin{align}
    \vec{\varphi}(x,t) & =  
    \hat{\phi}_{1}(x,t\mod N_{t}) - \hat{\phi}_{1}(x,t - 1\mod N_{t}) \nonumber \, ,\\[2ex]
    \; &  t\in\{2,\ldots, N_{t} \} \,,
\end{align}
and a zero mode
\[
\vec{\varphi}_{2}(x) = 
\sum_{t = 0}^{N_{t} - 1}\hat{\phi}_{1}(x,t)\,,
\]
for each space slice. We use the notation $\hat\phi$ to denote a unit vector in link
space corresponding to the variable $\phi$.

We have $V$ variables, which
are linear independent, since we can express each space-like link by
\begin{align}
\label{InvBasis}
\hat{\phi}_{1}(x,t) = 
\frac{1}{N_{t}}\Big(
\vec{\varphi}_{2}(x)
 &+ \sum_{k = 1}^{(t - 1)\mod N_{t}}k\vec{\varphi}(x,k)\\[1ex]
 &- \sum_{l = 1}^{(-t)\mod N_{t}}l\vec{\varphi}(x,N_{t} - l) \Big)\,.\nonumber
\end{align}
We replace the remaining $N_x$ time-like links
with plaquette variables
\begin{align}
\vec{\varphi}(x,0) &= \hat{\phi}_{2}(x,0) + \hat{\phi}_{1}(x,1)\\[1ex]
& - \hat{\phi}_{2}(x + 1 \mod N_x, 0) -  \hat{\phi}_{1}(x,0)\,,\nonumber
\end{align}
for $x\in\{1,\ldots,N_x - 1\}$ and the zero mode
\[
\vec{\varphi}_{1}(0) = 
\sum_{x = 0}^{N_x - 1}\hat{\phi}_{2}(x,0)\,.
\]
Using the fact that we already
have a basis transform for the space-like links,
we can express the remaining time-like links
in terms of these variables
in a similar fashion to Eq.~(\ref{InvBasis}).
This proves that our variable transformation
is invertible and therefore we have a 
non-zero Jacobian. By the
parametrization (\ref{LinkParam}) and the 
toron action (\ref{ToronAction}), we rewrite the
partition sum to
\begin{align}
 & Z \; = \; \int\mathrm{d}\theta_{1}(0)
\prod_x\mathrm{d}\theta_{2}(x)
\prod_{(x,t)\neq (0,0)}\mathrm{d}\theta(x,t)
\det\left[\frac{\partial \vec{\phi}}{\partial \vec{\theta}}\right] \nonumber \\[2ex]
& \; \times \; \exp\left[\beta/2 \sum_{(x,t)\neq (0,0)}\left(e^{i\theta(x,t)} + e^{-i\theta(x,t)}\right)\right]\\[3ex]
& \; \times \;
\exp\left[\beta/2\left(e^{-i\sum_{(x,t)\neq (0,0)} \theta(x,t)} + e^{i\sum_{(x,t)\neq (0,0)} \theta(x,t)}\right)
\right]\,.  \nonumber
\end{align}
Since our transformation is linear, the Jacobian
$\det\left[\frac{\partial \vec{\phi}}{\partial \vec{\theta}}\right]$
is constant and drops out when taking expectation
values. The same is true for the integrals over the
zero modes. They do not change the plaquettes and the  
 action. Their contribution is given by a constant
factor. Therefore they can be dropped without
changing expectation values. This reduces our degrees
of freedom by $N_x + 1$ to $V-1$.
The remaining effective degrees of freedom $\theta(x,t)$
are called
\emph{toron} variables.

\section{A local update procedure for a 2-dimensional \texorpdfstring{$\UOne$}{u(1)} gauge theory}
\label{AltUpdatesU1}
We parametrize
this tangent space by arclength. Therefore, we have
\begin{equation}
    \varphi_{\mu}(x) = \sqrt{\frac{\beta^*}{|\beta|}}
\lambda_{\mu}(x)\;,
\end{equation}
where $\lambda_{\mu}(x)\in\R$. 
We limit this space by the intersections seen in the one plaquette
model (see Sec.~\ref{sec:lefschetz-and-tang}). Therefore we have
\begin{align}
\nonumber
    \Lambda(x)&\;=\;\lambda_{0}(x)+\lambda_{1}(x+\hat{0})-
\lambda_{0}(x+\hat{1})-\lambda_{1}(x)
\\ & \; \in \; [-\pi R,+\pi R]\;, \quad \text{with}\quad
R = \mathrm{Re}\;\sqrt{\frac{\beta^*}{|\beta|}}\;.
\end{align}

We enforce this limit by setting the probability
for proposed configurations outside this limits to zero,
effectively rejecting them in the Metropolis step.
The algorithm is according to that

\begin{enumerate}
    \item Go through the lattice by a checker board pattern and pick
    accordingly $\lambda_{\mu}(x)$.
    \item Make a proposal 
    $\Delta \lambda_{\mu}(x) \sim \mathrm{Gauss}(\mu = 0, \sigma)$.
    and calculate the two adjacent plaquettes $\Lambda(x)$,
    $\Lambda(x-\hat{\nu})$ with $\nu\neq\mu$.
    \item If $\Lambda(x)$ and/or $\Lambda(x-\hat{\nu})$ is not within
    the boundaries, set $e^{-S'} = 0$. Otherwise calculate the 
    change of the action $\Delta S(P_{01}(x), P_{01}(x-\hat{\nu}))$
    \item Do a Metropolis Accept/Reject step and begin again from 1.
\end{enumerate}

Since our links can wander off into the imaginary direction,
we perform gauge cooling steps, see i.e. \cite{Seiler:2012wz}.
Locally for
a site $x$ we can analytically compute an optimum for the gauge
transformation $V(x)$. Therefore we look at the unitarity norm
\begin{align}
\label{UnitNorm}
    &\mathcal{F}[U_{\mu}(x)] \; = \\[1ex] \nonumber
    &\quad \sum_{x,\mu}\trace\left[
    U^{\dagger}_{\mu}(x)U_{\mu}(x) +
    (U^{\dagger}_{\mu}(x))^{-1}U^{-1}_{\mu}(x) - 2\mathbb{I}\right]\;,
\end{align}
which simplifies in our case $U_{\mu}(x)=e^{i\phi_{\mu}(x)}$ to
\begin{align}
\mathcal{F}[\phi_{\mu}(x)] & =
\sum_{x,\mu}\left[e^{-2\phi^I_{\mu}(x)} + e^{2\phi^I_{\mu}(x)}
- 2\right]\nonumber \\ \; & = \sum_{x,\mu}\left[|U_{\mu}(x)|^{-2} +
|U_{\mu}(x)|^2 - 2\right]\;,
\end{align}
depending only on the absolute value of the links. How a gauge transformation changes (\ref{UnitNorm}) depends therefore only
on its absolute value. For a local $V(x)$ we get the local minimum
at
\begin{equation}
    |V(x)| = \sqrt[4]{
    \frac{\sum_{\mu}
    \left( |U_{\mu}(x)|^{-2} + |U_{\mu}(x-\hat\mu)|^2\right)}
    {\sum_{\mu}
    \left(|U_{\mu}(x)|^{2} + |U_{\mu}(x-\hat\mu)|^{-2}\right)}},
    \label{LocGaugeCooling}
\end{equation}
which we apply between sweeps.\\
Alternatively one doesn't need to record the link values
and we restrict ourselves to just recording the plaquette values,
which are naturally bounded. We apply the update steps only in
terms of changes in links affecting their neighboring
plaquettes.\\
Reweighting to different orders of TMs can also be applied here
by injectively mapping the plaquette configurations to link configurations
and aligning the TMs bases to the unit basis, corresponding to
the individual links.
Given a set of plaquette values $\theta(x,t)$ in the algebra,
a possible mapping to link variables $\phi_{\mu}(x,t)$ would
be
\begin{enumerate}
    \item For $x=0$ and all $t\in 0,\ldots,N_{t}-1$ set
    \[
    \phi_2(0,t) = \theta(0,t).
    \]
    \item For $x=1$, all $t$, set
    \[\phi_2(2,t)=-\theta(1,t).\]
    \item Successively for $x\in 2,\ldots,N_x - 2$ and all $t$, set
    \[
    \phi_2(x+1,t) = \phi_2(x,t) - \theta(x,t).
    \]
    \item The last space-slice $x = N_x - 1$ is assigned
    successively by setting the first space-like link
    $\phi_1(x,0) = 0$ and for $t\in 1,\ldots,N_t - 1$
    to 
    \[\phi_1(x,t) = \phi_1(x,t-1) + \theta(x,t-1)
    + \phi_2(0,t-1) - \phi_2(x,t-1).
    \]
\end{enumerate}
The last plaquette value $\theta(N_x-1, N_t-1)$ is
automatically taken into account up to a factor of $2\pi$
by the periodic boundary conditions. Now the values obtained
for the algebra of the links are used to map onto the
subleading TMs.

\section{Proof of the matrix identity}
\label{midentity}
We consider the system of equations
\begin{equation}
\label{eq:mtgen}
    \trace[MT^a] = 0 \;\forall a,
\end{equation}
where $M$ is a general $N\times N$ complex matrix and
$T^a$ are generators of the lie algebra $\mathfrak{su}(N)$ or
$\mathfrak{u}(N)$. We show that Eq.~(\ref{eq:mtgen}) holds,
iff $M=c\mathbbm{1}$ for an arbitrary $c\in\C$ for
$\mathfrak{su}(N)$ and $M=0$ for $\mathfrak{u}(N)$.\\
First note, that the $T^a$ also form the basis of the
complex lie algebras $\mathfrak{sl}(N,\C)$,
$\mathfrak{gl}(N,\C)$. This is done by allowing complex
coefficients effectively removing the anti-hermiticity
of the elements. Consequently elements are only
defined by being traceless for $\mathfrak{sl}(N,\C)$
and we have $\mathfrak{gl}(N,\C)\simeq \C^{N\times N}$.\\
The statement does not depend on the
basis of the Lie algebra: Let $T^a$ and $S^a$ be two basis
of the same Lie algebra. We show that
\begin{equation}
    \label{basisind}
    \trace[MT^a] = 0 \;\forall a\;
    \Leftrightarrow\;
    \trace[MS^a] = 0 \;\forall a\;
\end{equation}
by expanding $S^a = \sum_b c^a_{b}T^b$ and using the
linearity of the trace
\begin{equation}
    \trace[MS^a] = \sum_b c^a_{b}\trace[MT^b] = 0.
\end{equation}
Especially the statement is equivalent to
\begin{equation}
\label{genstatement}
    \trace[MT] = 0 \;\forall T\in\mathfrak{sl}(N,\C)\;
    \text{or}\;
    \forall T\in\mathfrak{gl}(N,\C).
\end{equation}
Note, that the backward direction is trivial, since
$\trace[c\mathbbm{1}T]=c\trace[T]=0$ for $T$ being
traceless and $\trace[0T]=\trace[0]=0$ for $T\in\mathfrak{gl}(N,\C)$.\\
For the forward direction we use the more general
statement (\ref{genstatement}). For $T\in\mathfrak{sl}(N,\C)$, 
suppose first $M_{ij}\neq 0$ for $i\neq j$. Then we can choose
$T_{lk} = \delta_{lj}\delta_{ki}$, which is obviously traceless
and have
\begin{align}
    \trace[MT] = & \sum_{k,l}M_{kl}T_{lk} =
\sum_{k,l}M_{kl}\delta_{lj}\delta_{ki} \nonumber \\
= & M_{ij} \neq 0.
\end{align}
For $M_{ii}\neq M_{jj},\; i\neq j$ we take 
$T_{lk}=\delta_{li}\delta_{ki} - \delta_{lj}\delta_{kj}$,
which is also traceless and have
\begin{align}
    \trace[MT] = & \sum_{k,l}M_{kl}T_{lk} =
\sum_{k,l}M_{kl}
(\delta_{li}\delta_{ki} - \delta_{lj}\delta_{kj}) \nonumber \\
= &  M_{ii} - M_{jj} \neq 0.
\end{align}
Since $i$ and $j$ were arbitrary $M=c\mathbbm{1}$ for some
$c\in\C$.\\
For $T\in\mathfrak{gl}(N,\C)\sup \mathfrak{sl}(N,\C)$, $T$ is not traceless anymore and can also be e.g. $\mathbbm{1}$, which excludes
the $M=c\mathbbm{1}$ case and $M$ has to be zero to fulfill the
statement.

\bibliography{U1bib}{}
\end{document}